\documentclass[%
reprint,
superscriptaddress,
 amsmath,amssymb,
 aps,
 prb,
]{revtex4-2}

\usepackage{graphicx}% Include figure files
\usepackage{dcolumn}% Align table columns on decimal point
\usepackage{bm}% bold math

\usepackage{color}
\usepackage{ulem}

\begin{document}

%\preprint{APS/123-QED}

\title{Surface-termination-dependent electronic states in kagome superconductors $A\textrm{V}_{3}\textrm{Sb}_{5}$ ($A =$ K, Rb, Cs) studied by micro-ARPES}% Force line breaks with \\

\author{Takemi Kato}
\thanks{These authors contributed equally to this work.}
\affiliation{Department of Physics, Graduate School of Science, Tohoku University, Sendai 980-8578, Japan}

\author{Yongkai Li}
\thanks{These authors contributed equally to this work.}
\affiliation{Centre for Quantum Physics, Key Laboratory of Advanced Optoelectronic Quantum Architecture and Measurement (MOE), School of Physics, Beijing Institute of Technology, Beijing 100081, P. R. China}
\affiliation{Beijing Key Lab of Nanophotonics and Ultrafine Optoelectronic Systems, Beijing Institute of Technology, Beijing 100081, P. R. China}
\affiliation{Material Science Center, Yangtze Delta Region Academy of Beijing Institute of Technology, Jiaxing, 314011, P. R. China}

\author{Min Liu}
\thanks{These authors contributed equally to this work.}
\affiliation{Centre for Quantum Physics, Key Laboratory of Advanced Optoelectronic Quantum Architecture and Measurement (MOE), School of Physics, Beijing Institute of Technology, Beijing 100081, P. R. China}
\affiliation{Beijing Key Lab of Nanophotonics and Ultrafine Optoelectronic Systems, Beijing Institute of Technology, Beijing 100081, P. R. China}

\author{Kosuke Nakayama}
\thanks{Corresponding authors:\\
k.nakayama@arpes.phys.tohoku.ac.jp\\
zhiweiwang@bit.edu.cn\\
t-sato@arpes.phys.tohoku.ac.jp}
\affiliation{Department of Physics, Graduate School of Science, Tohoku University, Sendai 980-8578, Japan}
\affiliation{Precursory Research for Embryonic Science and Technology (PRESTO), Japan Science and Technology Agency (JST), Tokyo, 102-0076, Japan}

\author{Zhiwei Wang}
\thanks{Corresponding authors:\\
k.nakayama@arpes.phys.tohoku.ac.jp\\
zhiweiwang@bit.edu.cn\\
t-sato@arpes.phys.tohoku.ac.jp}
\affiliation{Centre for Quantum Physics, Key Laboratory of Advanced Optoelectronic Quantum Architecture and Measurement (MOE), School of Physics, Beijing Institute of Technology, Beijing 100081, P. R. China}
\affiliation{Beijing Key Lab of Nanophotonics and Ultrafine Optoelectronic Systems, Beijing Institute of Technology, Beijing 100081, P. R. China}
\affiliation{Material Science Center, Yangtze Delta Region Academy of Beijing Institute of Technology, Jiaxing, 314011, P. R. China}

\author{Seigo Souma}
\affiliation{Center for Science and Innovation in Spintronics (CSIS), Tohoku University, Sendai 980-8577, Japan}
\affiliation{Advanced Institute for Materials Research (WPI-AIMR), Tohoku University, Sendai 980-8577, Japan}

\author{Miho Kitamura}
\affiliation{Institute of Materials Structure Science, High Energy Accelerator Research Organization (KEK), Tsukuba, Ibaraki 305-0801, Japan}

\author{Koji Horiba}
\affiliation{Institute of Materials Structure Science, High Energy Accelerator Research Organization (KEK), Tsukuba, Ibaraki 305-0801, Japan}
\affiliation{National Institutes for Quantum Science and Technology (QST), Sendai 980-8579, Japan}

\author{Hiroshi Kumigashira}
\affiliation{Institute of Multidisciplinary Research for Advanced Materials (IMRAM), Tohoku University, Sendai 980-8577, Japan}

\author{Takashi Takahashi}
\affiliation{Department of Physics, Graduate School of Science, Tohoku University, Sendai 980-8578, Japan}

\author{Yugui Yao}
\affiliation{Centre for Quantum Physics, Key Laboratory of Advanced Optoelectronic Quantum Architecture and Measurement (MOE), School of Physics, Beijing Institute of Technology, Beijing 100081, P. R. China}
\affiliation{Beijing Key Lab of Nanophotonics and Ultrafine Optoelectronic Systems, Beijing Institute of Technology, Beijing 100081, P. R. China}

\author{Takafumi Sato}
\thanks{Corresponding authors:\\
k.nakayama@arpes.phys.tohoku.ac.jp\\
zhiweiwang@bit.edu.cn\\
t-sato@arpes.phys.tohoku.ac.jp}
\affiliation{Department of Physics, Graduate School of Science, Tohoku University, Sendai 980-8578, Japan}
\affiliation{Center for Science and Innovation in Spintronics (CSIS), Tohoku University, Sendai 980-8577, Japan}
\affiliation{Advanced Institute for Materials Research (WPI-AIMR), Tohoku University, Sendai 980-8577, Japan}
\affiliation{International Center for Synchrotron Radiation Innovation Smart (SRIS), Tohoku University, Sendai 980-8577, Japan}

\date{\today}

\begin{abstract}
Recently discovered kagome superconductors $A\textrm{V}_{3}\textrm{Sb}_{5}$ ($A =$ K, Rb, Cs) exhibit exotic bulk and surface physical properties such as charge-density wave (CDW) and chirality, whereas their origins remain unresolved.
By using micro-focused angle-resolved photoemission spectroscopy, we discovered that $A\textrm{V}_{3}\textrm{Sb}_{5}$ commonly exhibits two distinct polar surfaces depending on the termination; electron- and hole-doped ones for the $A$- and Sb-termination, respectively.
We observed that the kagome-derived band shows a clear splitting in the $A$-terminated surface while it is absent in the Sb-terminated counterpart, indicative of the polarity-dependent CDW at the surface.
Close comparison of the band-dependent splitting reveals that the three-dimensional CDW structure of the K-terminated surface is different from that of the Rb- or Cs-terminated surface, suggesting the diversity of the CDW ground state.
These results provide important insight into the origin of CDW in kagome superconductors $A\textrm{V}_{3}\textrm{Sb}_{5}$.
\end{abstract}

%\keywords{Suggested keywords}%Use showkeys class option if keyword
                              %display desired
\maketitle

\section{Introduction}
Kagome lattices, consisting of a two-dimensional network of corner-sharing triangles, offer a fertile platform to explore a variety of quantum phases, such as magnetic Weyl semimetals, unconventional density waves, charge fractionalization, and superconductivity \cite{TangPRL2011,SunPRL2011,NeupertPRL2011,WangPRL2011, YangNJP2017,LiuScience2019,MoraliScience2019,LiuNP2018,KurodaNM2017,NayakSciAdv2016,YinNP2019,LinPRL2018,YuPRB2012,WangPRB2013,KieselPRB2012,KieselPRL2013}, owing to the unique band structure due to the symmetry of the kagome lattice; flat band, Dirac cone, and saddle point forming a van Hove singularity.
Recent discovery of a new family of kagome metals $A\textrm{V}_{3}\textrm{Sb}_{5}$ ($A =$ K, Rb, Cs) which have a saddle point in the vicinity of the Fermi level ($E_{\textrm{F}}$) provides an excellent opportunity to investigate the charge-density wave (CDW; $T_{\textrm{CDW}}$ = 78--103 K) in the kagome lattice \cite{OrtizPRM2019,OrtizPRM2021,OrtizPRL2020,YinCPL2021}, which is accompanied by the three-dimensional (3D) $2\times2\times2$ or $2\times2\times4$ charge order \cite{LiangPRX2021,LiPRX2021,OrtizPRX2021}.
Besides CDW, $A\textrm{V}_{3}\textrm{Sb}_{5}$ commonly exhibits superconductivity ($T_{\textrm{c}}$ = 0.92--2.5 K) in the CDW phase \cite{OrtizPRM2019,OrtizPRM2021,OrtizPRL2020,YinCPL2021}.
The complex electronic phase diagram as a function of pressure and chemical doping indicates the intertwined nature of CDW and superconductivity \cite{DuPRB2021, ChenPRL2021, ZhangPRB2021, DuCPB2021, NakayamaPRX2022, SongPRL2021, OeyPRM2022, OeyPRM2022_2, LiuarXiv2021, YangSB2022, QianPRB2021, KatoPRL2022,LiuPRB2022, LiPRB2022}.
Emergence of other exotic properties, such as broken time reversal symmetry \cite{JiangNM2021, MielkeNature2022, ShumiyaPRB2021, WangPRB2021} and electronic nematicity \cite{NieNature2022, XuNP2022, JiangarXiv2022}, has been also reported in the CDW phase.
Thus, elucidating the origin of CDW and its interplay with other physical properties is urgently required to understand the exotic properties of $A\textrm{V}_{3}\textrm{Sb}_{5}$.

Besides unique bulk properties, $A\textrm{V}_{3}\textrm{Sb}_{5}$ shows several intriguing surface properties distinct from those of bulk, such as the unidirectional $4\times1$ stripe charge order and the surface-dependent vortex core state \cite{ LiangPRX2021, ShumiyaPRB2021, WangPRB2021, NieNature2022, ChenNature2021, ZhaoNature2021, YuNanoLett2022}.
These surface properties are strongly coupled to the crystal termination at the surface.
Namely, since $A\textrm{V}_{3}\textrm{Sb}_{5}$ consists of two building blocks, $A$ layer and $\textrm{V}_{3}\textrm{Sb}_{5}$ layer [Fig. 1(a)], cleavage of crystal produces either $A$- or Sb-terminated surface [Fig. 1(b)].
As demonstrated by scanning tunneling microscopy (STM) \cite{ LiangPRX2021, ShumiyaPRB2021, WangPRB2021, NieNature2022, ChenNature2021, ZhaoNature2021, YuNanoLett2022}, the $4\times1$ charge order and the vortex core states show a remarkable surface-termination dependence.
It was also uncovered by a recent angle-resolved photoemission spectroscopy (ARPES) with micro-focused photon beam that $\textrm{CsV}_{3}\textrm{Sb}_{5}$ shows a termination-dependent surface polarity that leads to the electron- and hole-rich characters for the Cs- and Sb-terminated surfaces, respectively \cite{KatoPRB2022}.
The ARPES study also revealed the electronic reconstruction due to the 3D CDW at the Cs-terminated surface and its suppression at the Sb-terminated surface \cite{KatoPRB2022}, indicating that the crystal termination modifies not only surface properties but also bulk-originated properties at the surface.
Therefore, high-spatial-resolution ARPES is needed to elucidate the bulk and surface band structures and their relationship with the physical properties in $A\textrm{V}_{3}\textrm{Sb}_{5}$.
However, such a study is scarce especially regarding the $A$-element dependence, despite the fact that the bulk CDW and surface stripe charge order show a strong $A$-element dependence \cite{LiangPRX2021, JiangNM2021, ShumiyaPRB2021, WangPRB2021, ChenNature2021, ZhaoNature2021, YuNanoLett2022, KangNM2022, LiNP2022, KautzscharXiv2022}.

In this paper, we report a spatially-resolved ARPES study of $\textrm{KV}_{3}\textrm{Sb}_{5}$, $\textrm{RbV}_{3}\textrm{Sb}_{5}$, and $\textrm{CsV}_{3}\textrm{Sb}_{5}$.
We systematically investigated the band structure of the $A$- and Sb-terminated surfaces of these materials, and found the universality of termination-dependent self-doping effect indicative of surface polarity.
We also found the universal and $A$-element dependent characteristics of 3D CDW.
We discuss implications of the present results in relation to the nature of exotic CDW and surface properties of $A\textrm{V}_{3}\textrm{Sb}_{5}$.

\begin{figure*}[htbp]
\includegraphics[width=178mm]{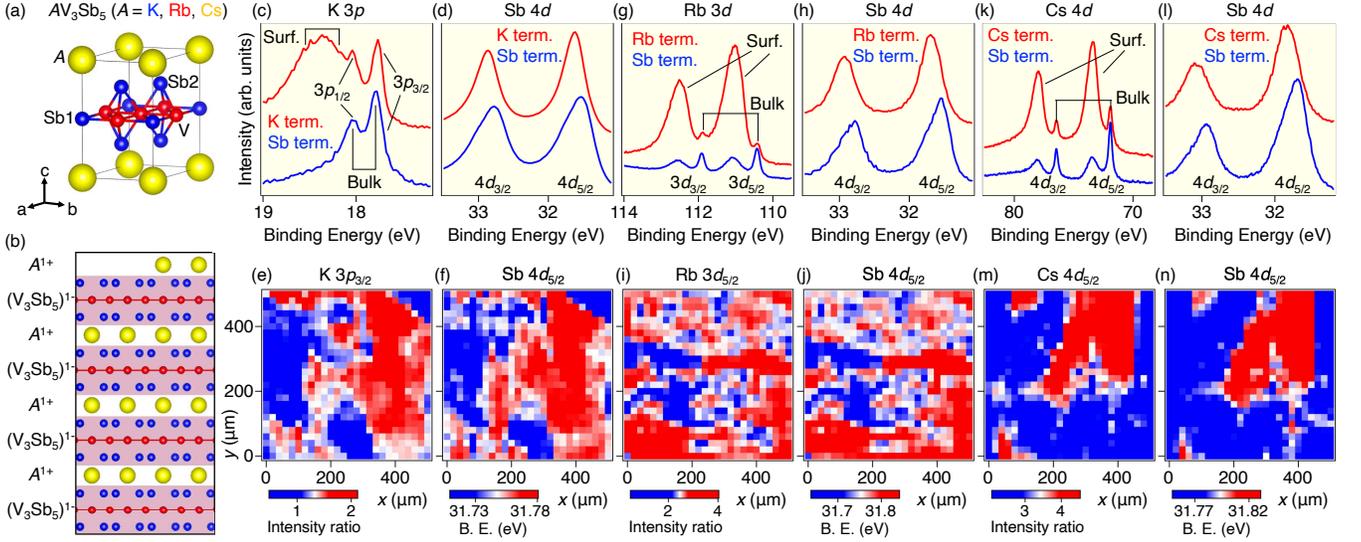}
 \caption{\label{Fig1}
 (a) Crystal structure of $A\textrm{V}_{3}\textrm{Sb}_{5}$ ($A =$ K, Rb, Cs).
 (b) Side view of crystal consisting of $A^{1+}$ and $(\textrm{V}_{3}\textrm{Sb}_{5})^{1-}$ layer.
 The cleaved surface is terminated by Sb or $A$ atoms.
 (c) and (d) EDCs in the K-$3p$ and Sb-$4d$ core-level regions, respectively, for the K-(red) and Sb-(blue) terminated surfaces of $\textrm{KV}_{3}\textrm{Sb}_{5}$.
 (e) and (f) Spatial map of the intensity ratio of the surface/bulk K-$3p_{3/2}$ core-level peaks and the $E_{\textrm{B}}$ position of Sb-$4d_{5/2}$ core level, respectively, measured at $T =$ 8 K in a surface area of $500\times500 \;\mu\textrm{m}^{2}$ for $\textrm{KV}_{3}\textrm{Sb}_{5}$.
 (g) and (h) EDCs in the Rb-$3d$ and Sb-$4d$ core-level regions, respectively, for the Rb-(red) and Sb-(blue) terminated surfaces of $\textrm{RbV}_{3}\textrm{Sb}_{5}$.
 (i) and (j) Same as (e) and (f) but for the Rb-$3d_{5/2}$ and Sb-$4d_{5/2}$ core-level peaks of $\textrm{RbV}_{3}\textrm{Sb}_{5}$, respectively.
 (k) and (l) EDCs in the Cs-$4d$ and Sb-$4d$ core-level regions, respectively, for the Cs-(red) and Sb-(blue) terminated surfaces of $\textrm{CsV}_{3}\textrm{Sb}_{5}$.
 (m) and (n) Same as (e) and (f) but for the Cs-$4d_{5/2}$ and Sb-$4d_{5/2}$ core-level peaks of $\textrm{CsV}_{3}\textrm{Sb}_{5}$, respectively.
 The core-level spectra were measured with $h\nu =$ 114, 150, and 106 eV for $\textrm{KV}_{3}\textrm{Sb}_{5}$, $\textrm{RbV}_{3}\textrm{Sb}_{5}$, and $\textrm{CsV}_{3}\textrm{Sb}_{5}$, respectively. }
 \end{figure*}
 
\section{Experiments}
\subsection{Crystal growth and ARPES measurements}
High-quality $A\textrm{V}_{3}\textrm{Sb}_{5}$ single crystals ($A =$ K, Rb, Cs) were synthesized by the self-flux method \cite{WangPRB2021}.
ARPES measurements were performed using Scienta-Omicron DA30 and SES2002 spectrometers at BL-28A and BL-2A in Photon Factory, KEK, with the micro-focused beam (spot size of $10\times12 \:\mu\textrm{m}^{2}$) achieved by Kirkpatrick-Baez mirror optics \cite{KitamuraRSI2022}.
We used energy tunable photons of $h\nu =$ 85--350 eV.
ARPES data were mainly obtained by circularly polarized 114-, 111-, and 106-eV photons for $\textrm{KV}_{3}\textrm{Sb}_{5}$, $\textrm{RbV}_{3}\textrm{Sb}_{5}$, and $\textrm{CsV}_{3}\textrm{Sb}_{5}$, respectively (corresponding to the out-of-plane wave vector $k_{z}\sim0$; see Appendix A).
The energy resolution was set to be 25 meV.
Samples were quickly cooled down from room temperature to $T = 8$ K, and then cleaved and measured.
ARPES measurements were also carried out using soft x ray with Scienta SES2002 spectrometer at BL-2A in Photon Factory, KEK.

\subsection{Band calculations}
The first-principles calculations were performed based on the density functional theory (DFT), as implemented in the Vienna Ab initio Simulation Package (VASP) \cite{KressePRB1996, KresseCMS1996}.
The projected augmented wave (PAW) is adopted as the pseudopotentials \cite{BlochlPRB1994}.
The generalized gradient approximation with the Perdew-Burke-Ernzerhof (PBE) realization was used for the exchange-correlation functional \cite{PerdewPRL1996}.
We used an energy cutoff of 500 eV and a $12\times12\times7$ $\Gamma$-centered k-points for the self-consistent calculations.
The interlayer van der Waals interactions are taken into account using the DFT-D3 correction \cite{GrimmeJCP2010}.
The lattice constants and internal coordinates are optimized until the atomic forces become less than $10^{-4}$ eV/$\textrm{\AA}$.
We used the WANNIERTOOLS package \cite{WuCPC2018} to perform the slab band calculations, in which we construct a 3-unit-cell slab structure with the Sb ($A$) termination at the top (bottom) surface.

\section{Results and Discussion}
First, to investigate the surface termination, we measured spatially-resolved core level spectra of $\textrm{KV}_{3}\textrm{Sb}_{5}$ [Figs. 1(c)--1(f)], $\textrm{RbV}_{3}\textrm{Sb}_{5}$ [Figs. 1(g)--1(j)], and $\textrm{CsV}_{3}\textrm{Sb}_{5}$ [Figs. 1(k)--1(n)] by sweeping the micro-focused beam in the area of $500\times500 \;\mu\textrm{m}^{2}$ of cleaved surfaces.
According to a previous study on $\textrm{CsV}_{3}\textrm{Sb}_{5}$ \cite{KatoPRB2022}, the intensity ratio between the surface and bulk Cs-$4d$ core peaks due to Cs atoms at the topmost surface and beneath the $\textrm{V}_{3}\textrm{Sb}_{5}$ layer, respectively, is a good measure for determining the surface termination.
As seen from representative Cs-$4d$ core-level spectra in Fig. 1(k), each spin-orbit satellite (Cs $4d_{5/2}$ or $4d_{3/2}$) consists of surface and bulk components located at higher and lower binding energies ($E_{\textrm{B}}$'s), respectively.
The Cs-termination-dominant surface (red curve) is characterized by a stronger surface-derived peak than the bulk one, whereas the Sb-termination-dominant surface (blue curve) is characterized by a stronger bulk-derived peak \cite{KatoPRB2022}.
As displayed in Fig. 1(m), the spatial dependence of the surface/bulk Cs-$4d$ core-peak intensity-ratio clearly maps out the Cs- and Sb-termination-dominant domains (red and blue regions, respectively).
Hereafter, we simply call them Cs- and Sb-terminated surfaces, respectively.
An intriguing property in relation to the difference in the surface termination is polarity \cite{KatoPRB2022}.
This is seen in the Sb-$4d$ core-level spectra in Fig. 1(l), in which both the Sb-$4d_{5/2}$ and $4d_{3/2}$ peaks of the Cs-terminated surface (red curve) shift to higher $E_{\textrm{B}}$ by $\sim$130 meV compared to those of the Sb-terminated one (blue curve) due to the polarity-induced self-electron-doping.
Polar domains visualized by the spatial map at the $E_{\textrm{B}}$ value of the Sb-$4d_{5/2}$ core-level peak [Fig. 1(n)] show almost the same spatial distribution as that in Fig. 1(m), indicating a close link between the surface polarity and the termination.
In the same way, we have mapped out the surface termination and polarity for $\textrm{KV}_{3}\textrm{Sb}_{5}$ and $\textrm{RbV}_{3}\textrm{Sb}_{5}$, and found an overall similarity to those observed for $\textrm{CsV}_{3}\textrm{Sb}_{5}$.
As seen from Figs. 1(c) and 1(g), the K-$3p$ and Rb-$3d$ core levels consist of the bulk and surface components as in the case of $\textrm{CsV}_{3}\textrm{Sb}_{5}$ (note that the surface K-$3p_{3/2}$ and $3p_{1/2}$ peaks appear as a single broad peak due to their small energy separation).
The stronger intensity of the surface component compared to the bulk one in the red curve signifies the $A$ termination, whereas the weaker intensity of the surface component in the blue curve suggests the Sb termination.
In the Sb-$4d$ core level, the peak position for the $A$ termination [red curves in Figs. 1(d) and 1(h)] shifts to higher $E_{\textrm{B}}$ by 100--150 meV compared to that for the Sb termination (blue curves), indicating the polar-surface formation.
Figures 1(e) and 1(i), and Figs. 1(f) and 1(j) show the spatial distribution of the surface termination and polarity, respectively.
All the images include the red and blue regions which correspond to the electron-doped K/Rb- and hole-doped Sb-rich surfaces, respectively.
In addition, the red- and blue-colored patterns are almost identical between Figs. 1(e) and 1(f), and Figs. 1(i) and 1(j), showing the universality of the termination-dependent surface polarity in $A\textrm{V}_{3}\textrm{Sb}_{5}$ (note that the domain size appeared to be different at each cleavage and position of surface, suggesting that the $A$-dependent variation in the domain size seen in Fig. 1 may not be an essential feature).

\begin{figure}[htbp]
\includegraphics[width=86mm]{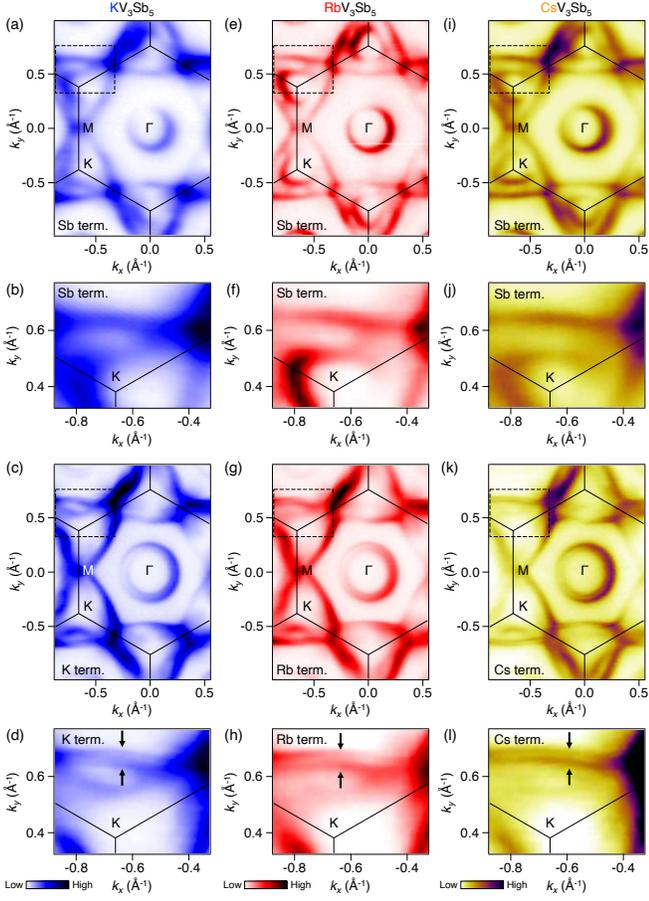}
 \caption{\label{Fig2}
(a) and (b) Fermi-surface mapping for the Sb-terminated surface of $\textrm{KV}_{3}\textrm{Sb}_{5}$ measured at $T =$ 8 K and an enlarged view in the $\textbf{k}$ region enclosed by a dashed rectangle in (a), respectively.
(c) and (d) Same as (a) and (b), respectively, but for the K-terminated surface.
(e)--(h) Same as (a)--(d) but for $\textrm{RbV}_{3}\textrm{Sb}_{5}$.
(i)--(l) Same as (a)--(d) but for $\textrm{CsV}_{3}\textrm{Sb}_{5}$.}
\end{figure}

The universal feature among $A\textrm{V}_{3}\textrm{Sb}_{5}$ is also seen near $E_{\textrm{F}}$.
Figure 2 shows the Fermi-surface (FS) mapping measured at $T = 8$ K (well below bulk $T_{\textrm{CDW}}$) for $\textrm{KV}_{3}\textrm{Sb}_{5}$ [Figs. 2(a)--2(d)], $\textrm{RbV}_{3}\textrm{Sb}_{5}$ [Figs. 2(e)--2(h)], and $\textrm{CsV}_{3}\textrm{Sb}_{5}$ [Figs. 2(i)--2(l)].
We have selected $k_{z}\sim0$ plane for the mapping by sweeping photon energies and thereby tracing the $k_{z}$ dispersion (see Appendix A for details).
In the Sb-terminated surface [Figs. 2(a), 2(e), and 2(i)], the FS of all $A\textrm{V}_{3}\textrm{Sb}_{5}$ consists of a circular pocket centered at the $\Gamma$ point, a slightly distorted hexagon near the Brillouin-zone boundary, and a triangular pocket centered at the K point, in qualitative agreement with first-principles bulk-band calculations in the normal state \cite{JiangNM2021, ZhaoNature2021, KangNM2022, NakayamaPRB2021, WangPRB2022, TanPRL2021, FuPRL2021, KangNP2022, HuNCOM2022, LuoNCOM2022, LiuPRX2021, ChoPRL2021, LiPRR2022, KatoCOMMAT2022}.
The circular pocket is attributed to the 5$p_{z}$ band of Sb atoms embedded in the kagome-lattice plane [Sb1 in Fig. 1(a)], while the hexagonal and triangular FSs are mainly due to the V kagome-derived $3d_{xz}/d_{yz}$ and $3d_{xy}/d_{x^{2}-y^{2}}$ bands, respectively.
In the $A$-terminated surface [Figs. 2(c), 2(g), and 2(k)], one can find that each of the circular and hexagonal FSs consists of two sheets [see two concentric circles centered at the $\Gamma$ point in Figs. 2(c), 2(g), and 2(k) and two parallel lines indicated by arrows in Figs. 2(d), 2(h), and 2(l)], in contrast to their single-sheet nature in the Sb-terminated surface [Figs. 2(a), 2(b), 2(e), 2(f), 2(i), and 2(j)].
In addition to the increase in the number of FS, one can find an expansion of the circular pockets in the $A$-terminated surface compared to that in the Sb-terminated surface (see Appendix B for the comparison of FS volume).
These results demonstrate that $A\textrm{V}_{3}\textrm{Sb}_{5}$ commonly shows termination-dependent FSs.

To understand the termination-dependent electronic states in more detail, we next measured the band dispersions near $E_{\textrm{F}}$.
We plot the ARPES intensity obtained for the Sb-terminated surface at $T = 8$ K along the KMK cut ($k_{x} = -\pi$) in Figs. 3(a)--3(c) for $\textrm{KV}_{3}\textrm{Sb}_{5}$, $\textrm{RbV}_{3}\textrm{Sb}_{5}$, and $\textrm{CsV}_{3}\textrm{Sb}_{5}$, respectively.
In all $A\textrm{V}_{3}\textrm{Sb}_{5}$, there are highly dispersive $\beta$ and $\gamma$ bands crossing $E_{\textrm{F}}$ midway between the $\Gamma$ and K points, the $\delta$ and $\varepsilon$ bands forming a Dirac cone at the K point, and the hole-like $\eta$ band topped at the M point.
In the $A$-terminated surface [Figs. 3(d)--3(f)], two key differences compared to the Sb-terminated surfaces are commonly observed regardless of the $A$ element.
Firstly, several bands in the $A$-terminated surface are shifted downward compared to those in the Sb-terminated one.
For instance, the Dirac point in the $A$-terminated surface shifts downward by $\sim$30--60 meV compared to that in the Sb-terminated one (compare black arrows; also see Appendix C for detailed comparison).
Also, the top of the $\delta$ band shifts downward to approach $E_{\textrm{F}}$, so that it shows a bending behavior near $E_{\textrm{F}}$.
These band shifts are consistent with the polarity-induced doping.
It is noted that the bending behavior of the $\delta$ band is partly associated with the termination-dependent CDW \cite{KatoPRB2022}, as discussed later.
Secondly, several V-derived bands exhibit splitting in the $A$-terminated surface.
For instance, the $\beta$ band splits into two subbands, as corroborated by comparison of momentum distribution curves (MDCs) near $E_{\textrm{F}}$ between the $A$- and Sb-terminated surfaces [Figs. 3(j), 3(l), and 3(n); see the double-vs-single peaked feature in MDCs for the $A$- and Sb-termination, respectively].
This splitting accounts for the presence of two hexagonal FSs in Figs. 2(d), 2(h), and 2(l).
In addition, the $\delta$ and $\eta$ bands split into two subbands in the $A$-terminated surface.
The splitting of $\delta$ band is supported by two-peaked structure in the energy distribution curve (EDC) near the Fermi wave vector [Figs. 3(k), 3(m), and 3(o)] and the splitting of the $\eta$ band is seen in the second-derivative intensity plot [Figs. 3(g), 3(h), and 3(i)].
The band dispersion near $E_{\textrm{F}}$ is strongly termination-dependent irrespective of the $A$ element.

\begin{figure}[htbp]
\includegraphics[width=86mm]{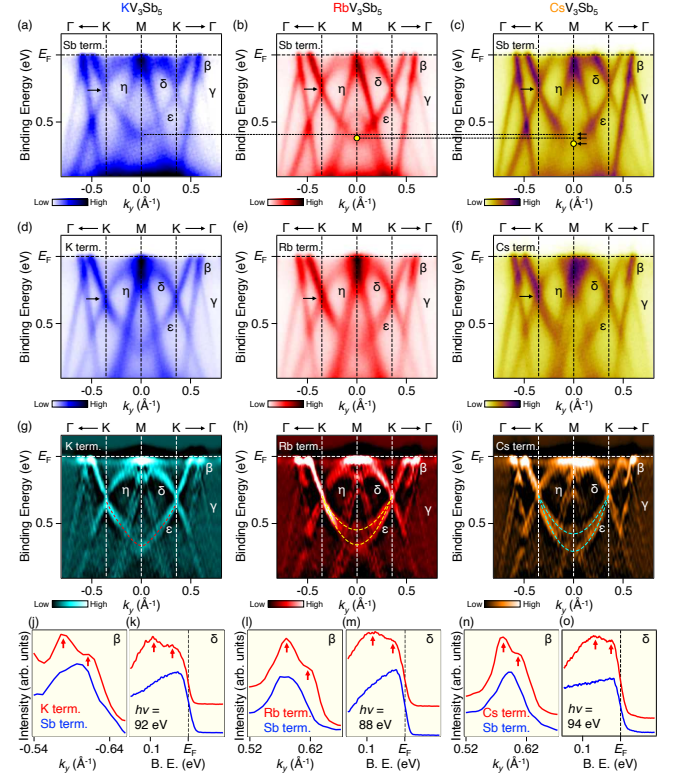}
 \caption{\label{Fig3}
(a)--(c) ARPES intensity plots at $T =$ 8 K measured along $k_{y}$ at $k_{x} = -\pi$ (KMK cut) for Sb-terminated surface of $\textrm{KV}_{3}\textrm{Sb}_{5}$, $\textrm{RbV}_{3}\textrm{Sb}_{5}$, and $\textrm{CsV}_{3}\textrm{Sb}_{5}$, respectively.
(d)--(f) Same as (a)--(c) but for K-, Rb-, and Cs-terminated surface, respectively.
(g)--(i) Second-derivative intensity of (d)--(f), respectively.
(j) Comparison of MDC of the $\beta$ band at $E_{\textrm{F}}$ between the K- and Sb-terminated surfaces.
(k) Comparison of EDC of the $\delta$ band near the $k_{\textrm{F}}$ point between the two surface terminations.
(l) and (m) Same as (j) and (k) but for $\textrm{RbV}_{3}\textrm{Sb}_{5}$.
(n) and (o) Same as (j) and (k) but for $\textrm{CsV}_{3}\textrm{Sb}_{5}$.
Red arrows in (j)--(o) indicate the splitting of the $\beta$ and $\delta$ bands in the K-, Rb-, and Cs-terminated surfaces.}
\end{figure}

Besides the universality among $A\textrm{V}_{3}\textrm{Sb}_{5}$ demonstrated above, we found an $A$-element dependence of the electronic states.
Side-by-side comparison of the band dispersions in the Sb-terminated surface [Figs. 3(a)--3(c)] shows that the bottom of the $\varepsilon$ band at the M point monotonically shifts downward from $\textrm{KV}_{3}\textrm{Sb}_{5}$ to $\textrm{CsV}_{3}\textrm{Sb}_{5}$, as indicated by yellow circles and black dashed lines in Figs. 3(a)--3(c).
The $\varepsilon$ band also exhibits $A$-dependent dispersion in the $A$-terminated surface.
As seen from Figs. 3(g)--3(i), the $\varepsilon$ band splits into two subbands around the M point in $\textrm{RbV}_{3}\textrm{Sb}_{5}$ and $\textrm{CsV}_{3}\textrm{Sb}_{5}$ [Figs. 3(h) and 3(i)], whereas it is not clearly visible in $\textrm{KV}_{3}\textrm{Sb}_{5}$ [Fig. 3(g)], in sharp contrast to the $\beta$, $\gamma$, and $\eta$ bands in which the splitting is commonly observed in all $A\textrm{V}_{3}\textrm{Sb}_{5}$.

The $A$-element dependence of electronic states is also observed around the $\Gamma$ point. 
Figures 4(a)--4(c) and 4(d)--4(f) show the ARPES intensity along the K$\Gamma$K cut for the Sb-terminated surface and the corresponding second-derivative intensity, respectively.
There are two electron pockets ($\alpha$ and $\alpha$' bands) which produce circular FSs as shown in Fig. 2.
The inner $\alpha$ band is the bulk band because it shows a clear $k_{z}$ dispersion (see Appendix A in the $k_{z}$-dependent ARPES), while the outer $\alpha$' band is possibly of surface origin \cite{LiPRR2022}.
One can find that, while the position of $\alpha$' band is not sensitive to the $A$ element, the bottom of $\alpha$ band shows a clear $A$-dependent energy shift; $E_{\textrm{B}} = 0.60$ eV for $\textrm{KV}_{3}\textrm{Sb}_{5}$, 0.57 eV for $\textrm{RbV}_{3}\textrm{Sb}_{5}$, and 0.53 eV for $\textrm{CsV}_{3}\textrm{Sb}_{5}$.
The systematic energy shift of the $\alpha$-band bottom is also observed in the Sb-terminated surface [see Figs. 4(g)--4(i)].

\begin{figure}[htbp]
\includegraphics[width=86mm]{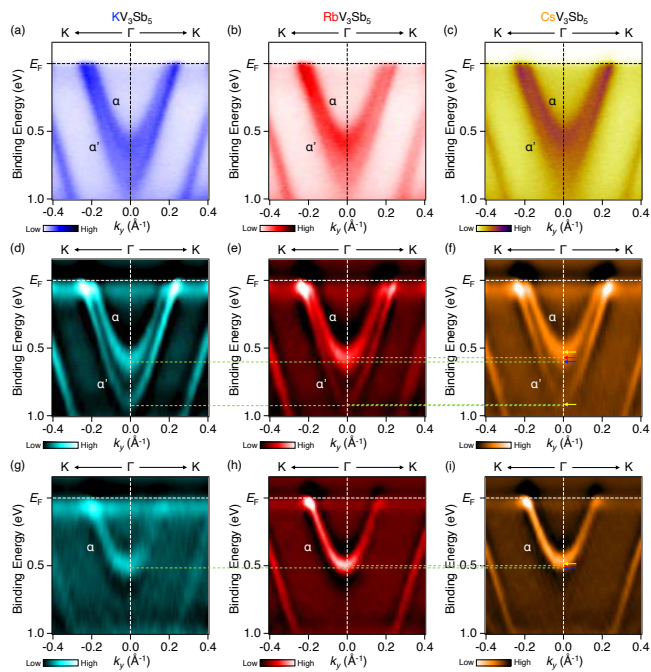}
 \caption{\label{Fig4}
(a)--(c) ARPES intensity plots at $T =$ 8 K measured around the $\Gamma$ point for K-, Rb-, and Cs-terminated surfaces of $\textrm{KV}_{3}\textrm{Sb}_{5}$, $\textrm{RbV}_{3}\textrm{Sb}_{5}$, and $\textrm{CsV}_{3}\textrm{Sb}_{5}$, respectively.
(d)--(f) Second-derivative intensity plots of (a)--(c), respectively.
Dashed lines and arrows indicate the bottom of $\alpha$  and $\alpha$' bands.
(g)--(i) Same as (d)--(f) but for the Sb-terminated surface.}
 \end{figure}
 
To explain the observed surface-termination and $A$-element dependences of the band structure, we performed first-principles band structure calculations for a 3-unit-cell slab with the Sb and $A$ terminations at the top and bottom surfaces, respectively [Fig. 5(a)].
Figures 5(b)--5(d) are the calculated band structure along the $\Gamma$KM cut, in which red, gray, and blue colorings represent the contribution from the Sb-terminated surface at the top, the bulk inside the slab, and the $A$-terminated surface at the bottom, respectively.
One can notice that the $\alpha$ and $\delta$ bands of the Sb-terminated surface shift upward compared to the $A$-terminated counterparts.
This reasonably agrees with the experimental observation, supporting the polar nature of the termination-dependent band structure.
It is noted that, although the atomic rearrangement often modifies the band dispersion at the surface, we assumed the undistorted $1\times1$ lattice in the present calculations, so the main trigger of the termination-dependent electronic states is unlikely atomic rearrangement but a band-bending effect due to the electrostatic potential gradient, i.e., polarity.
The modulation of band structure and carrier concentration at the surface also indicates a need for caution when discussing/comparing physical properties probed by surface- and bulk-sensitive probes in $A\textrm{V}_{3}\textrm{Sb}_{5}$.
Besides the surface-termination dependence, some $A$-element dependences in the band structure are also captured by our calculations.
As indicated by dashed lines and arrows in Fig. 5(d), the bottom of $\alpha$ band gradually shifts downward in going from $\textrm{KV}_{3}\textrm{Sb}_{5}$, to $\textrm{RbV}_{3}\textrm{Sb}_{5}$ and $\textrm{CsV}_{3}\textrm{Sb}_{5}$.
Also, the bottom of $\varepsilon$ band at the M point shows a similar tendency in qualitative agreement with the experimental observation.
The $A$-dependent band shift likely originates from a variation of the relative energy position and the width of band due to a difference in the lattice parameter.

\begin{figure*}[htbp]
\includegraphics[width=140mm]{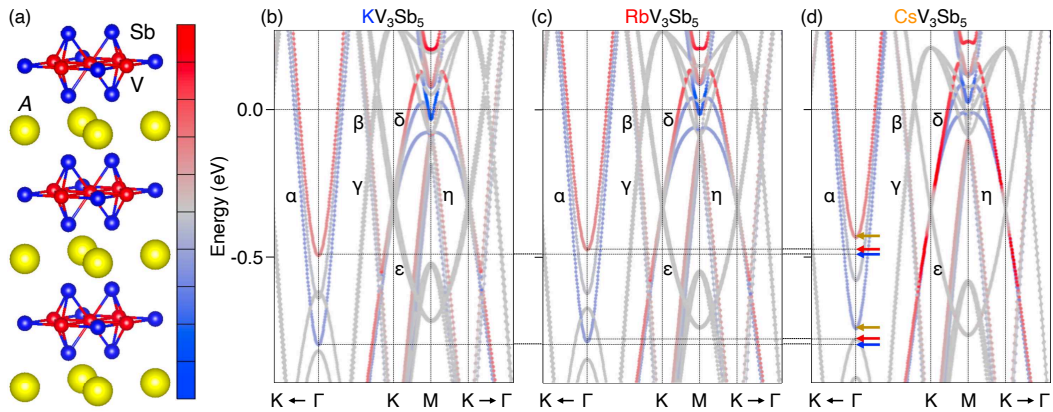}
 \caption{\label{Fig5}
(a) Slab structure for which first-principles band calculations were performed.
(b)--(d) Calculated band structures along the $\Gamma$KM cut for $\textrm{KV}_{3}\textrm{Sb}_{5}$, $\textrm{RbV}_{3}\textrm{Sb}_{5}$, and $\textrm{CsV}_{3}\textrm{Sb}_{5}$ slabs, respectively.
Dashed lines and arrows indicate the bottom of $\alpha$ band.
Red and blue dots represent the contribution from the top and bottom surfaces, as indicated by gradual shading in (a).}
 \end{figure*}
 
 \begin{table*}[htbp]
\caption{\label{Tab1}%
Summary of $A$-element and surface-termination dependences of band structure and 3D CDW.
$E_{\textrm{bottom}}$ (eV) indicates the binding energy of the band bottom. }
\begin{ruledtabular}
\begin{tabular}{l|>{\centering}p{18mm}c|>{\centering}p{26mm}c|>{\centering}p{26mm}c}
 {}& \multicolumn{2}{c}{$\textrm{KV}_{3}\textrm{Sb}_{5}$} & \multicolumn{2}{c}{$\textrm{RbV}_{3}\textrm{Sb}_{5}$} & \multicolumn{2}{c}{$\textrm{CsV}_{3}\textrm{Sb}_{5}$}\\
\hline
Surface termination&K rich&Sb rich&Rb rich&Sb rich&Cs rich&Sb rich\\
$E_{\textrm{bottom}}$ of $\alpha$ band&0.60&0.50&0.57&0.49&0.53&0.46\\
$E_{\textrm{bottom}}$ of $\alpha$' band&0.89&---&0.88&---&0.88&---\\
Doped carrier&electron&hole&electron&hole&electron&hole\\
Split of $\beta$, $\gamma$, $\delta$ bands&yes&no&yes&no&yes&no\\
Split of $\varepsilon$ band&no&no&yes&no&yes&no\\
3D CDW&Staggered SoD or TrH&no&Alternate stacking of SoD and TrH&no&Alternate stacking of SoD and TrH&no\\
\end{tabular}
\end{ruledtabular}
\end{table*}

Now we discuss implications of the present results in relation to CDW.
As demonstrated above, the slab calculation with the $1\times1$ structure shows a reasonable agreement with the experimental band structure in the Sb-terminated surface, whereas the calculation cannot reproduce the band splitting in the $A$-terminated surface.
Since the splitting of the V-$3d$ band is probably due to the out-of-plane unit-cell doubling and the resultant band folding along $k_{z}$ \cite{KatoPRB2022, KangNM2022, LiPRR2022, HuPRB2022}, the present result strongly suggests the formation of $2\times2\times2$ CDW in the $A$-terminated surface and its suppression in the Sb-terminated one.
While such termination/polarity-dependent CDW was suggested in a previous ARPES on $\textrm{CsV}_{3}\textrm{Sb}_{5}$ \cite{KatoPRB2022}, the present result has clearly established the universality among $A\textrm{V}_{3}\textrm{Sb}_{5}$.

Present results also provide deeper insight into the nature of CDW.
According to the model calculations for various types of 3D CDW \cite{KangNM2022, HuPRB2022}, the $\varepsilon$ band splits when the kagome layers with the star-of-David (SoD)-type and tri-hexagonal (TrH)-type distortions are stacked alternately, whereas it does not split when the kagome layers have only SoD or TrH distortion.
Therefore, the present observations of the absence of $\varepsilon$-band splitting in $\textrm{KV}_{3}\textrm{Sb}_{5}$ and its presence in $\textrm{RbV}_{3}\textrm{Sb}_{5}$ and $\textrm{CsV}_{3}\textrm{Sb}_{5}$ suggest that the $2\times2\times2$ CDW in $\textrm{KV}_{3}\textrm{Sb}_{5}$ originates from only SoD or TrH distortion with a $\pi$ phase shift along the $c$ axis (staggered SoD or TrH CDW), while the $2\times2\times2$ CDW in $\textrm{RbV}_{3}\textrm{Sb}_{5}$ and $\textrm{CsV}_{3}\textrm{Sb}_{5}$ is an alternate stacking of the SoD and TrH distortions \cite{KangNM2022, HuPRB2022} (see Table 1 for summary of the present key results).
In this regard, it is noted that, while we observed the $\varepsilon$-band splitting in $\textrm{RbV}_{3}\textrm{Sb}_{5}$, it was absent in a recent study by another group \cite{KangNM2022}.
This difference implies that the type of 3D CDW at the surface is not robust even in the same material.
Possible key factors determining the type of 3D CDW are carrier/chemical doping, cooling speed, etc. \cite{KatoPRB2022, KangNM2022, XiaoarXiv2022}.

Another exotic property of $A\textrm{V}_{3}\textrm{Sb}_{5}$ is the unidirectional $4\times1$ charge order that appears only in the Sb-terminated surface.
According to STM studies, the $4\times1$ charge order is present in $\textrm{RbV}_{3}\textrm{Sb}_{5}$ and $\textrm{CsV}_{3}\textrm{Sb}_{5}$ but absent in $\textrm{KV}_{3}\textrm{Sb}_{5}$ \cite{LiangPRX2021, JiangNM2021, ShumiyaPRB2021, WangPRB2021, ChenNature2021, ZhaoNature2021, YuNanoLett2022, LiNCOM2022, LiarXiv2022}.
Despite such $A$-element dependence of the presence/absence of the $4\times1$ charge order, the present result shows that the FS topology and size in the Sb-terminated surface are similar among $A\textrm{V}_{3}\textrm{Sb}_{5}$ and a good FS nesting condition with the $4\times1$ periodicity is not observed in any of $A\textrm{V}_{3}\textrm{Sb}_{5}$, suggesting that the mechanism of the unidirectional order may be beyond the framework with simple FS nesting.

In conclusion, we reported a systematic study on the spatially-resolved band structure of $A\textrm{V}_{3}\textrm{Sb}_{5}$ by micro-ARPES in combination with first-principles band structure calculations.
We have established the universality of the polar nature of cleaved surfaces characterized by the opposite band energy shift between the $A$- and Sb-terminated surfaces.
We also found that the presence/absence and the microscopic structure of 3D CDW depend on the surface polarity (termination) and the $A$ element, pointing to an exceptional sensitivity of 3D CDW to the underlying electronic and chemical environments.
The present results provide foundation for understanding and manipulating exotic physical properties of $A\textrm{V}_{3}\textrm{Sb}_{5}$.

\begin{acknowledgments}
This work was supported by JST-CREST (No. JPMJCR18T1), JST-PRESTO (No. JPMJPR18L7), Grant-in-Aid for Scientific Research (JSPS KAKENHI Grant Numbers JP21H04435 and JP20H01847), KEK-PF (Proposal number: 2021S2-001), and the Sasakawa Scientific Research Grant from the Japan Science Society.
The work at Beijing Institute of Technology was supported by the National Key R\&D Program of China (Grant Nos. 2020YFA0308800, 2022YFA1403400), the Natural Science Foundation of China (Grant No. 92065109), and the Beijing Natural Science Foundation (Grant Nos. Z210006, Z190006).
T.K. acknowledges support from GP-Spin at Tohoku University and JST-SPRING (No. JPMJSP2114).
Z.W. thanks the Analysis \& Testing Center at BIT for assistance in facility support.
\end{acknowledgments}

\section*{APPENDIX A: Normal emission measurement}
Figure 6 shows the second-derivative ARPES intensity measured at the normal-emission set up in the Rb-terminated surface of $\textrm{RbV}_{3}\textrm{Sb}_{5}$.
The energy position of the inner electron-band ($\alpha$) bottom shows a periodic dispersion as a function of $k_{z}$ (see red dashed curve; note that the intensity of the outer electron band is weak), suggesting the bulk-band nature.
From the observed periodicity, we have estimated the inner-potential value to be $V_{\textrm{0}} = 10.0$ eV, which indicates that the $k_{z}\sim0$ plane is probed by 111-eV photons in $\textrm{RbV}_{3}\textrm{Sb}_{5}$.
In $\textrm{KV}_{3}\textrm{Sb}_{5}$ and $\textrm{CsV}_{3}\textrm{Sb}_{5}$, the $k_{z}\sim0$ plane is probed by $h\nu =$ 114 eV and 106 eV, respectively, according to our previous works \cite{NakayamaPRB2021,KatoCOMMAT2022}.

 \begin{figure}[htbp]
\includegraphics[width=46mm]{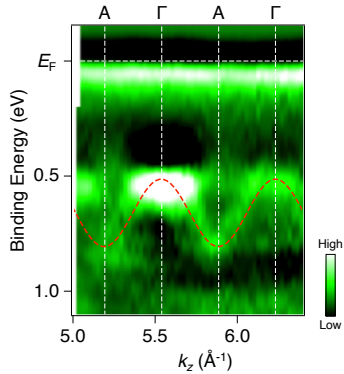}
 \caption{\label{Fig6}
Second-derivative intensity of the ARPES spectra for $\textrm{RbV}_{3}\textrm{Sb}_{5}$ plotted as a function of $k_{z}$ and binding energy obtained with photon energies of 90--150 eV.}
 \end{figure}
 
\section*{APPENDIX B: Surface-termination dependence of the Fermi-surface size}
Figure 7 shows side-by-side comparison of the ARPES intensity plots at $E_{\textrm{F}}$ obtained in the Sb- and $A$-terminated surfaces of $\textrm{KV}_{3}\textrm{Sb}_{5}$ [Figs. 7(a) and (b)], $\textrm{RbV}_{3}\textrm{Sb}_{5}$ [Figs. 7(c) and 7(d)], and $\textrm{CsV}_{3}\textrm{Sb}_{5}$ [Figs. 7(e) and 7(f)] (note that the data are identical to those in Fig. 2).
One can see that both the two circular FSs centered at the $\Gamma$ point in the $A$-terminated surface are larger than the one in the Sb-terminated surface, in agreement with the electron-doped character of the $A$-terminated surface.

 \begin{figure}[htbp]
\includegraphics[width=86mm]{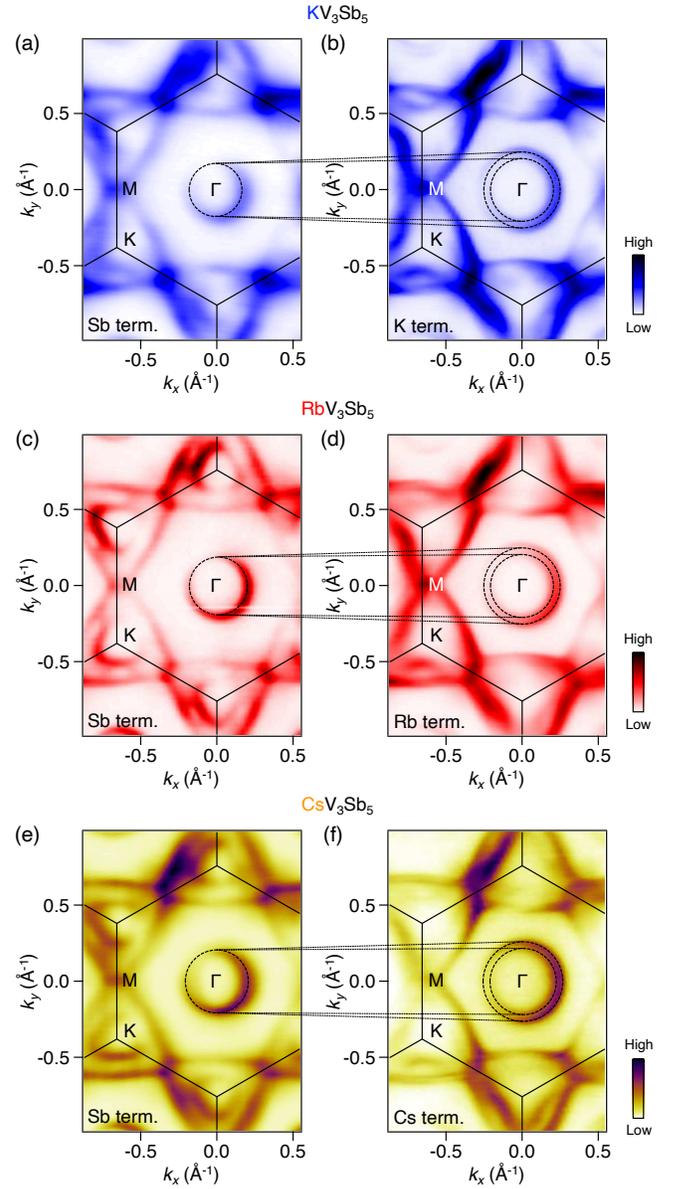}
 \caption{\label{Fig7}
 (a) and (b) Fermi-surface mapping in the Sb- and $A$-terminated surfaces, respectively, in $\textrm{KV}_{3}\textrm{Sb}_{5}$.
 (c) and (d) Same as (a) and (b) but in $\textrm{RbV}_{3}\textrm{Sb}_{5}$.
 (e) and (f) Same as (a) and (b) but measured in $\textrm{CsV}_{3}\textrm{Sb}_{5}$.}
 \end{figure}

\section*{APPENDIX C: Comparison of the band dispersions between Sb- and A-terminated surfaces}
Figure 8 shows comparison of the band dispersions along the KMK cut between the Sb- and $A$-terminated surfaces of $\textrm{KV}_{3}\textrm{Sb}_{5}$ [Figs. 8(a) and 8(b)], $\textrm{RbV}_{3}\textrm{Sb}_{5}$ [Figs. 8(c) and 8(d)], and $\textrm{CsV}_{3}\textrm{Sb}_{5}$ [Figs. 8(e) and 8(f)].
The data are identical to those in Fig. 3, but the energy shift of the Dirac point is better recognized by side-by-side comparison.
One can also see that the $\delta$ band near $E_{\textrm{F}}$ in the $A$-terminated surface is shifted downward compared to that in the Sb-terminated surface (dotted line).

 \begin{figure}[htbp]
\includegraphics[width=86mm]{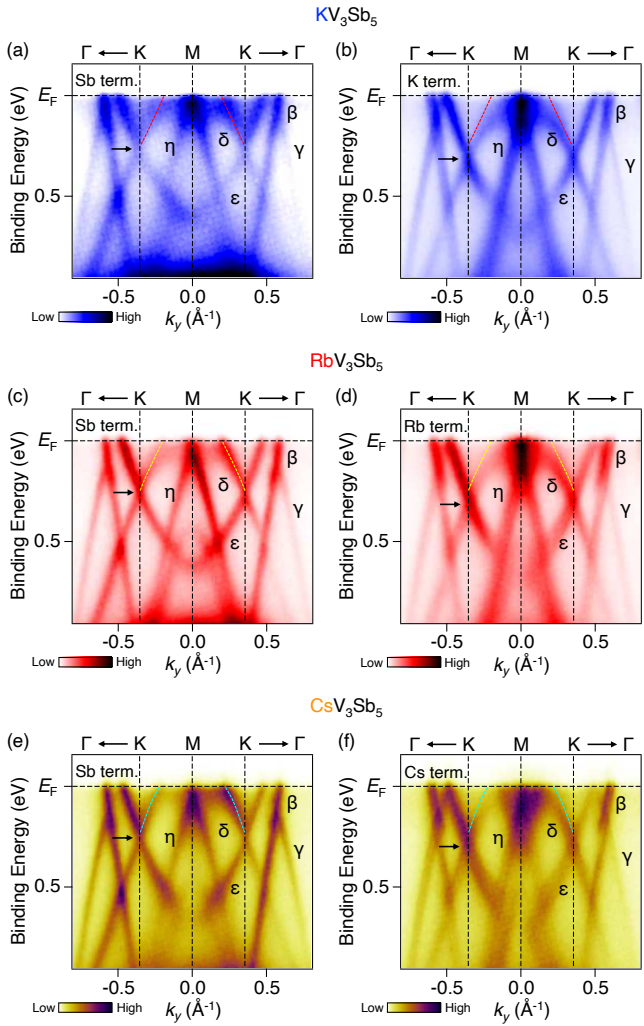}
 \caption{\label{Fig8}
(a) and (b) ARPES intensity plots at $T =$ 8 K measured along $k_{y}$ at $k_{x} = -\pi$ (KMK cut) for the Sb- and $A$-terminated surfaces, respectively, in $\textrm{KV}_{3}\textrm{Sb}_{5}$.
(c) and (d) Same as (a) and (b), respectively, but measured in $\textrm{RbV}_{3}\textrm{Sb}_{5}$.
(e) and (f) Same as (a) and (b), respectively, but measured in $\textrm{CsV}_{3}\textrm{Sb}_{5}$.
Black arrows indicate the energy position of the Dirac point at the K point.
Dotted lines are the $\delta$-band dispersion determined in the Sb-terminated surface.}
 \end{figure}

%\bibliography{AVStermination}% Produces the bibliography via BibTeX.

\begin{thebibliography}{73}%
\makeatletter
\providecommand \@ifxundefined [1]{%
 \@ifx{#1\undefined}
}%
\providecommand \@ifnum [1]{%
 \ifnum #1\expandafter \@firstoftwo
 \else \expandafter \@secondoftwo
 \fi
}%
\providecommand \@ifx [1]{%
 \ifx #1\expandafter \@firstoftwo
 \else \expandafter \@secondoftwo
 \fi
}%
\providecommand \natexlab [1]{#1}%
\providecommand \enquote  [1]{``#1''}%
\providecommand \bibnamefont  [1]{#1}%
\providecommand \bibfnamefont [1]{#1}%
\providecommand \citenamefont [1]{#1}%
\providecommand \href@noop [0]{\@secondoftwo}%
\providecommand \href [0]{\begingroup \@sanitize@url \@href}%
\providecommand \@href[1]{\@@startlink{#1}\@@href}%
\providecommand \@@href[1]{\endgroup#1\@@endlink}%
\providecommand \@sanitize@url [0]{\catcode `\\12\catcode `\$12\catcode
  `\&12\catcode `\#12\catcode `\^12\catcode `\_12\catcode `\%12\relax}%
\providecommand \@@startlink[1]{}%
\providecommand \@@endlink[0]{}%
\providecommand \url  [0]{\begingroup\@sanitize@url \@url }%
\providecommand \@url [1]{\endgroup\@href {#1}{\urlprefix }}%
\providecommand \urlprefix  [0]{URL }%
\providecommand \Eprint [0]{\href }%
\providecommand \doibase [0]{https://doi.org/}%
\providecommand \selectlanguage [0]{\@gobble}%
\providecommand \bibinfo  [0]{\@secondoftwo}%
\providecommand \bibfield  [0]{\@secondoftwo}%
\providecommand \translation [1]{[#1]}%
\providecommand \BibitemOpen [0]{}%
\providecommand \bibitemStop [0]{}%
\providecommand \bibitemNoStop [0]{.\EOS\space}%
\providecommand \EOS [0]{\spacefactor3000\relax}%
\providecommand \BibitemShut  [1]{\csname bibitem#1\endcsname}%
\let\auto@bib@innerbib\@empty
%</preamble>
\bibitem [{\citenamefont {Tang}\ \emph {et~al.}(2011)\citenamefont {Tang},
  \citenamefont {Mei},\ and\ \citenamefont {Wen}}]{TangPRL2011}%
  \BibitemOpen
  \bibfield  {author} {\bibinfo {author} {\bibfnamefont {E.}~\bibnamefont
  {Tang}}, \bibinfo {author} {\bibfnamefont {J.-W.}\ \bibnamefont {Mei}},\ and\
  \bibinfo {author} {\bibfnamefont {X.-G.}\ \bibnamefont {Wen}},\ }\href@noop
  {} {\bibfield  {journal} {\bibinfo  {journal} {Phys.\ Rev.\ Lett.}\ }\textbf
  {\bibinfo {volume} {106}},\ \bibinfo {pages} {236802} (\bibinfo {year}
  {2011})}\BibitemShut {NoStop}%
\bibitem [{\citenamefont {Sun}\ \emph {et~al.}(2011)\citenamefont {Sun},
  \citenamefont {Gu}, \citenamefont {Katsura},\ and\ \citenamefont
  {Das~Sarma}}]{SunPRL2011}%
  \BibitemOpen
  \bibfield  {author} {\bibinfo {author} {\bibfnamefont {K.}~\bibnamefont
  {Sun}}, \bibinfo {author} {\bibfnamefont {Z.}~\bibnamefont {Gu}}, \bibinfo
  {author} {\bibfnamefont {H.}~\bibnamefont {Katsura}},\ and\ \bibinfo {author}
  {\bibfnamefont {S.}~\bibnamefont {Das~Sarma}},\ }\href@noop {} {\bibfield
  {journal} {\bibinfo  {journal} {Phys.\ Rev.\ Lett.}\ }\textbf {\bibinfo
  {volume} {106}},\ \bibinfo {pages} {236803} (\bibinfo {year}
  {2011})}\BibitemShut {NoStop}%
\bibitem [{\citenamefont {Neupert}\ \emph {et~al.}(2011)\citenamefont
  {Neupert}, \citenamefont {Santos}, \citenamefont {Chamon},\ and\
  \citenamefont {Mudry}}]{NeupertPRL2011}%
  \BibitemOpen
  \bibfield  {author} {\bibinfo {author} {\bibfnamefont {T.}~\bibnamefont
  {Neupert}}, \bibinfo {author} {\bibfnamefont {L.}~\bibnamefont {Santos}},
  \bibinfo {author} {\bibfnamefont {C.}~\bibnamefont {Chamon}},\ and\ \bibinfo
  {author} {\bibfnamefont {C.}~\bibnamefont {Mudry}},\ }\href@noop {}
  {\bibfield  {journal} {\bibinfo  {journal} {Phys.\ Rev.\ Lett.}\ }\textbf
  {\bibinfo {volume} {106}},\ \bibinfo {pages} {236804} (\bibinfo {year}
  {2011})}\BibitemShut {NoStop}%
\bibitem [{\citenamefont {Wang}\ \emph {et~al.}(2011)\citenamefont {Wang},
  \citenamefont {Gu}, \citenamefont {Gong},\ and\ \citenamefont
  {Sheng}}]{WangPRL2011}%
  \BibitemOpen
  \bibfield  {author} {\bibinfo {author} {\bibfnamefont {Y.-F.}\ \bibnamefont
  {Wang}}, \bibinfo {author} {\bibfnamefont {Z.-C.}\ \bibnamefont {Gu}},
  \bibinfo {author} {\bibfnamefont {C.-D.}\ \bibnamefont {Gong}},\ and\
  \bibinfo {author} {\bibfnamefont {D.~N.}\ \bibnamefont {Sheng}},\ }\href@noop
  {} {\bibfield  {journal} {\bibinfo  {journal} {Phys.\ Rev.\ Lett.}\ }\textbf
  {\bibinfo {volume} {107}},\ \bibinfo {pages} {146803} (\bibinfo {year}
  {2011})}\BibitemShut {NoStop}%
\bibitem [{\citenamefont {Yang}\ \emph {et~al.}(2017)\citenamefont {Yang},
  \citenamefont {Sun}, \citenamefont {Zhang}, \citenamefont {Shi},
  \citenamefont {Parkin},\ and\ \citenamefont {Yan}}]{YangNJP2017}%
  \BibitemOpen
  \bibfield  {author} {\bibinfo {author} {\bibfnamefont {H.}~\bibnamefont
  {Yang}}, \bibinfo {author} {\bibfnamefont {Y.}~\bibnamefont {Sun}}, \bibinfo
  {author} {\bibfnamefont {Y.}~\bibnamefont {Zhang}}, \bibinfo {author}
  {\bibfnamefont {W.-J.}\ \bibnamefont {Shi}}, \bibinfo {author} {\bibfnamefont
  {S.~S.~P.}\ \bibnamefont {Parkin}},\ and\ \bibinfo {author} {\bibfnamefont
  {B.}~\bibnamefont {Yan}},\ }\href@noop {} {\bibfield  {journal} {\bibinfo
  {journal} {New J. Phys.}\ }\textbf {\bibinfo {volume} {19}},\ \bibinfo
  {pages} {015008} (\bibinfo {year} {2017})}\BibitemShut {NoStop}%
\bibitem [{\citenamefont {Liu}\ \emph {et~al.}(2019)\citenamefont {Liu},
  \citenamefont {Liang}, \citenamefont {Liu}, \citenamefont {Xu}, \citenamefont
  {Li}, \citenamefont {Chen}, \citenamefont {Pei}, \citenamefont {Shi},
  \citenamefont {Mo}, \citenamefont {Dudin}, \citenamefont {Kim}, \citenamefont
  {Cacho}, \citenamefont {Li}, \citenamefont {Sun}, \citenamefont {Yang},
  \citenamefont {Liu}, \citenamefont {Parkin}, \citenamefont {Felser},\ and\
  \citenamefont {Chen}}]{LiuScience2019}%
  \BibitemOpen
  \bibfield  {author} {\bibinfo {author} {\bibfnamefont {D.~F.}\ \bibnamefont
  {Liu}}, \bibinfo {author} {\bibfnamefont {A.~J.}\ \bibnamefont {Liang}},
  \bibinfo {author} {\bibfnamefont {E.~K.}\ \bibnamefont {Liu}}, \bibinfo
  {author} {\bibfnamefont {Q.~N.}\ \bibnamefont {Xu}}, \bibinfo {author}
  {\bibfnamefont {Y.~W.}\ \bibnamefont {Li}}, \bibinfo {author} {\bibfnamefont
  {C.}~\bibnamefont {Chen}}, \bibinfo {author} {\bibfnamefont {D.}~\bibnamefont
  {Pei}}, \bibinfo {author} {\bibfnamefont {W.~J.}\ \bibnamefont {Shi}},
  \bibinfo {author} {\bibfnamefont {S.~K.}\ \bibnamefont {Mo}}, \bibinfo
  {author} {\bibfnamefont {P.}~\bibnamefont {Dudin}}, \bibinfo {author}
  {\bibfnamefont {T.}~\bibnamefont {Kim}}, \bibinfo {author} {\bibfnamefont
  {C.}~\bibnamefont {Cacho}}, \bibinfo {author} {\bibfnamefont
  {G.}~\bibnamefont {Li}}, \bibinfo {author} {\bibfnamefont {Y.}~\bibnamefont
  {Sun}}, \bibinfo {author} {\bibfnamefont {L.~X.}\ \bibnamefont {Yang}},
  \bibinfo {author} {\bibfnamefont {Z.~K.}\ \bibnamefont {Liu}}, \bibinfo
  {author} {\bibfnamefont {S.~S.~P.}\ \bibnamefont {Parkin}}, \bibinfo {author}
  {\bibfnamefont {C.}~\bibnamefont {Felser}},\ and\ \bibinfo {author}
  {\bibfnamefont {Y.~L.}\ \bibnamefont {Chen}},\ }\href@noop {} {\bibfield
  {journal} {\bibinfo  {journal} {Science}\ }\textbf {\bibinfo {volume}
  {365}},\ \bibinfo {pages} {1282} (\bibinfo {year} {2019})}\BibitemShut
  {NoStop}%
\bibitem [{\citenamefont {Morali}\ \emph {et~al.}(2019)\citenamefont {Morali},
  \citenamefont {Batabyal}, \citenamefont {Nag}, \citenamefont {Liu},
  \citenamefont {Xu}, \citenamefont {Sun}, \citenamefont {Yan}, \citenamefont
  {Felser}, \citenamefont {Avraham},\ and\ \citenamefont
  {Beidenkopf}}]{MoraliScience2019}%
  \BibitemOpen
  \bibfield  {author} {\bibinfo {author} {\bibfnamefont {N.}~\bibnamefont
  {Morali}}, \bibinfo {author} {\bibfnamefont {R.}~\bibnamefont {Batabyal}},
  \bibinfo {author} {\bibfnamefont {P.~K.}\ \bibnamefont {Nag}}, \bibinfo
  {author} {\bibfnamefont {E.}~\bibnamefont {Liu}}, \bibinfo {author}
  {\bibfnamefont {Q.}~\bibnamefont {Xu}}, \bibinfo {author} {\bibfnamefont
  {Y.}~\bibnamefont {Sun}}, \bibinfo {author} {\bibfnamefont {B.}~\bibnamefont
  {Yan}}, \bibinfo {author} {\bibfnamefont {C.}~\bibnamefont {Felser}},
  \bibinfo {author} {\bibfnamefont {N.}~\bibnamefont {Avraham}},\ and\ \bibinfo
  {author} {\bibfnamefont {H.}~\bibnamefont {Beidenkopf}},\ }\href@noop {}
  {\bibfield  {journal} {\bibinfo  {journal} {Science}\ }\textbf {\bibinfo
  {volume} {365}},\ \bibinfo {pages} {1286} (\bibinfo {year}
  {2019})}\BibitemShut {NoStop}%
\bibitem [{\citenamefont {Liu}\ \emph {et~al.}(2018)\citenamefont {Liu},
  \citenamefont {Sun}, \citenamefont {Kumar}, \citenamefont {Muechler},
  \citenamefont {Sun}, \citenamefont {Jiao}, \citenamefont {Yang},
  \citenamefont {Liu}, \citenamefont {Liang}, \citenamefont {Xu}, \citenamefont
  {Kroder}, \citenamefont {Süß}, \citenamefont {Borrmann}, \citenamefont
  {Shekhar}, \citenamefont {Wang}, \citenamefont {Xi}, \citenamefont {Wang},
  \citenamefont {Schnelle}, \citenamefont {Wirth}, \citenamefont {Chen},
  \citenamefont {Goennenwein},\ and\ \citenamefont {Felser}}]{LiuNP2018}%
  \BibitemOpen
  \bibfield  {author} {\bibinfo {author} {\bibfnamefont {E.}~\bibnamefont
  {Liu}}, \bibinfo {author} {\bibfnamefont {Y.}~\bibnamefont {Sun}}, \bibinfo
  {author} {\bibfnamefont {N.}~\bibnamefont {Kumar}}, \bibinfo {author}
  {\bibfnamefont {L.}~\bibnamefont {Muechler}}, \bibinfo {author}
  {\bibfnamefont {A.}~\bibnamefont {Sun}}, \bibinfo {author} {\bibfnamefont
  {L.}~\bibnamefont {Jiao}}, \bibinfo {author} {\bibfnamefont {S.-Y.}\
  \bibnamefont {Yang}}, \bibinfo {author} {\bibfnamefont {D.}~\bibnamefont
  {Liu}}, \bibinfo {author} {\bibfnamefont {A.}~\bibnamefont {Liang}}, \bibinfo
  {author} {\bibfnamefont {Q.}~\bibnamefont {Xu}}, \bibinfo {author}
  {\bibfnamefont {J.}~\bibnamefont {Kroder}}, \bibinfo {author} {\bibfnamefont
  {V.}~\bibnamefont {Süß}}, \bibinfo {author} {\bibfnamefont
  {H.}~\bibnamefont {Borrmann}}, \bibinfo {author} {\bibfnamefont
  {C.}~\bibnamefont {Shekhar}}, \bibinfo {author} {\bibfnamefont
  {Z.}~\bibnamefont {Wang}}, \bibinfo {author} {\bibfnamefont {C.}~\bibnamefont
  {Xi}}, \bibinfo {author} {\bibfnamefont {W.}~\bibnamefont {Wang}}, \bibinfo
  {author} {\bibfnamefont {W.}~\bibnamefont {Schnelle}}, \bibinfo {author}
  {\bibfnamefont {S.}~\bibnamefont {Wirth}}, \bibinfo {author} {\bibfnamefont
  {Y.}~\bibnamefont {Chen}}, \bibinfo {author} {\bibfnamefont {S.~T.~B.}\
  \bibnamefont {Goennenwein}},\ and\ \bibinfo {author} {\bibfnamefont
  {C.}~\bibnamefont {Felser}},\ }\href@noop {} {\bibfield  {journal} {\bibinfo
  {journal} {Nat. Phys.}\ }\textbf {\bibinfo {volume} {14}},\ \bibinfo {pages}
  {1125} (\bibinfo {year} {2018})}\BibitemShut {NoStop}%
\bibitem [{\citenamefont {Kuroda}\ \emph {et~al.}(2017)\citenamefont {Kuroda},
  \citenamefont {Tomita}, \citenamefont {Suzuki}, \citenamefont {Bareille},
  \citenamefont {Nugroho}, \citenamefont {Goswami}, \citenamefont {Ochi},
  \citenamefont {Ikhlas}, \citenamefont {Nakayama}, \citenamefont {Akebi},
  \citenamefont {Noguchi}, \citenamefont {Ishii}, \citenamefont {Inami},
  \citenamefont {Ono}, \citenamefont {Kumigashira}, \citenamefont {Varykhalov},
  \citenamefont {Muro}, \citenamefont {Koretsune}, \citenamefont {Arita},
  \citenamefont {Shin}, \citenamefont {Kondo},\ and\ \citenamefont
  {Nakatsuji}}]{KurodaNM2017}%
  \BibitemOpen
  \bibfield  {author} {\bibinfo {author} {\bibfnamefont {K.}~\bibnamefont
  {Kuroda}}, \bibinfo {author} {\bibfnamefont {T.}~\bibnamefont {Tomita}},
  \bibinfo {author} {\bibfnamefont {M.-T.}\ \bibnamefont {Suzuki}}, \bibinfo
  {author} {\bibfnamefont {C.}~\bibnamefont {Bareille}}, \bibinfo {author}
  {\bibfnamefont {A.~A.}\ \bibnamefont {Nugroho}}, \bibinfo {author}
  {\bibfnamefont {P.}~\bibnamefont {Goswami}}, \bibinfo {author} {\bibfnamefont
  {M.}~\bibnamefont {Ochi}}, \bibinfo {author} {\bibfnamefont {M.}~\bibnamefont
  {Ikhlas}}, \bibinfo {author} {\bibfnamefont {M.}~\bibnamefont {Nakayama}},
  \bibinfo {author} {\bibfnamefont {S.}~\bibnamefont {Akebi}}, \bibinfo
  {author} {\bibfnamefont {R.}~\bibnamefont {Noguchi}}, \bibinfo {author}
  {\bibfnamefont {R.}~\bibnamefont {Ishii}}, \bibinfo {author} {\bibfnamefont
  {N.}~\bibnamefont {Inami}}, \bibinfo {author} {\bibfnamefont
  {K.}~\bibnamefont {Ono}}, \bibinfo {author} {\bibfnamefont {H.}~\bibnamefont
  {Kumigashira}}, \bibinfo {author} {\bibfnamefont {A.}~\bibnamefont
  {Varykhalov}}, \bibinfo {author} {\bibfnamefont {T.}~\bibnamefont {Muro}},
  \bibinfo {author} {\bibfnamefont {T.}~\bibnamefont {Koretsune}}, \bibinfo
  {author} {\bibfnamefont {R.}~\bibnamefont {Arita}}, \bibinfo {author}
  {\bibfnamefont {S.}~\bibnamefont {Shin}}, \bibinfo {author} {\bibfnamefont
  {T.}~\bibnamefont {Kondo}},\ and\ \bibinfo {author} {\bibfnamefont
  {S.}~\bibnamefont {Nakatsuji}},\ }\href@noop {} {\bibfield  {journal}
  {\bibinfo  {journal} {Nat. Mater.}\ }\textbf {\bibinfo {volume} {16}},\
  \bibinfo {pages} {1090} (\bibinfo {year} {2017})}\BibitemShut {NoStop}%
\bibitem [{\citenamefont {Nayak}\ \emph {et~al.}(2016)\citenamefont {Nayak},
  \citenamefont {Fischer}, \citenamefont {Sun}, \citenamefont {Yan},
  \citenamefont {Karel}, \citenamefont {Komarek}, \citenamefont {Shekhar},
  \citenamefont {Kumar}, \citenamefont {Schnelle}, \citenamefont {Kübler},
  \citenamefont {Felser},\ and\ \citenamefont {Parkin}}]{NayakSciAdv2016}%
  \BibitemOpen
  \bibfield  {author} {\bibinfo {author} {\bibfnamefont {A.~K.}\ \bibnamefont
  {Nayak}}, \bibinfo {author} {\bibfnamefont {J.~E.}\ \bibnamefont {Fischer}},
  \bibinfo {author} {\bibfnamefont {Y.}~\bibnamefont {Sun}}, \bibinfo {author}
  {\bibfnamefont {B.}~\bibnamefont {Yan}}, \bibinfo {author} {\bibfnamefont
  {J.}~\bibnamefont {Karel}}, \bibinfo {author} {\bibfnamefont {A.~C.}\
  \bibnamefont {Komarek}}, \bibinfo {author} {\bibfnamefont {C.}~\bibnamefont
  {Shekhar}}, \bibinfo {author} {\bibfnamefont {N.}~\bibnamefont {Kumar}},
  \bibinfo {author} {\bibfnamefont {W.}~\bibnamefont {Schnelle}}, \bibinfo
  {author} {\bibfnamefont {J.}~\bibnamefont {Kübler}}, \bibinfo {author}
  {\bibfnamefont {C.}~\bibnamefont {Felser}},\ and\ \bibinfo {author}
  {\bibfnamefont {S.~S.~P.}\ \bibnamefont {Parkin}},\ }\href@noop {} {\bibfield
   {journal} {\bibinfo  {journal} {Sci. Adv.}\ }\textbf {\bibinfo {volume}
  {2}},\ \bibinfo {pages} {e1501870} (\bibinfo {year} {2016})}\BibitemShut
  {NoStop}%
\bibitem [{\citenamefont {Yin}\ \emph {et~al.}(2019)\citenamefont {Yin},
  \citenamefont {Zhang}, \citenamefont {Chang}, \citenamefont {Wang},
  \citenamefont {Tsirkin}, \citenamefont {Guguchia}, \citenamefont {Lian},
  \citenamefont {Zhou}, \citenamefont {Jiang}, \citenamefont {Belopolski},
  \citenamefont {Shumiya}, \citenamefont {Multer}, \citenamefont {Litskevich},
  \citenamefont {Cochran}, \citenamefont {Lin}, \citenamefont {Wang},
  \citenamefont {Neupert}, \citenamefont {Jia}, \citenamefont {Lei},\ and\
  \citenamefont {Hasan}}]{YinNP2019}%
  \BibitemOpen
  \bibfield  {author} {\bibinfo {author} {\bibfnamefont {J.-X.}\ \bibnamefont
  {Yin}}, \bibinfo {author} {\bibfnamefont {S.~S.}\ \bibnamefont {Zhang}},
  \bibinfo {author} {\bibfnamefont {G.}~\bibnamefont {Chang}}, \bibinfo
  {author} {\bibfnamefont {Q.}~\bibnamefont {Wang}}, \bibinfo {author}
  {\bibfnamefont {S.~S.}\ \bibnamefont {Tsirkin}}, \bibinfo {author}
  {\bibfnamefont {Z.}~\bibnamefont {Guguchia}}, \bibinfo {author}
  {\bibfnamefont {B.}~\bibnamefont {Lian}}, \bibinfo {author} {\bibfnamefont
  {H.}~\bibnamefont {Zhou}}, \bibinfo {author} {\bibfnamefont {K.}~\bibnamefont
  {Jiang}}, \bibinfo {author} {\bibfnamefont {I.}~\bibnamefont {Belopolski}},
  \bibinfo {author} {\bibfnamefont {N.}~\bibnamefont {Shumiya}}, \bibinfo
  {author} {\bibfnamefont {D.}~\bibnamefont {Multer}}, \bibinfo {author}
  {\bibfnamefont {M.}~\bibnamefont {Litskevich}}, \bibinfo {author}
  {\bibfnamefont {T.~A.}\ \bibnamefont {Cochran}}, \bibinfo {author}
  {\bibfnamefont {H.}~\bibnamefont {Lin}}, \bibinfo {author} {\bibfnamefont
  {Z.}~\bibnamefont {Wang}}, \bibinfo {author} {\bibfnamefont {T.}~\bibnamefont
  {Neupert}}, \bibinfo {author} {\bibfnamefont {S.}~\bibnamefont {Jia}},
  \bibinfo {author} {\bibfnamefont {H.}~\bibnamefont {Lei}},\ and\ \bibinfo
  {author} {\bibfnamefont {M.~Z.}\ \bibnamefont {Hasan}},\ }\href@noop {}
  {\bibfield  {journal} {\bibinfo  {journal} {Nat. Phys.}\ }\textbf {\bibinfo
  {volume} {15}},\ \bibinfo {pages} {443} (\bibinfo {year} {2019})}\BibitemShut
  {NoStop}%
\bibitem [{\citenamefont {Lin}\ \emph {et~al.}(2018)\citenamefont {Lin},
  \citenamefont {Choi}, \citenamefont {Zhang}, \citenamefont {Qin},
  \citenamefont {Yi}, \citenamefont {Wang}, \citenamefont {Li}, \citenamefont
  {Wang}, \citenamefont {Zhang}, \citenamefont {Sun}, \citenamefont {Wei},
  \citenamefont {Zhang}, \citenamefont {Guo}, \citenamefont {Lu}, \citenamefont
  {Cho}, \citenamefont {Zeng},\ and\ \citenamefont {Zhang}}]{LinPRL2018}%
  \BibitemOpen
  \bibfield  {author} {\bibinfo {author} {\bibfnamefont {Z.}~\bibnamefont
  {Lin}}, \bibinfo {author} {\bibfnamefont {J.-H.}\ \bibnamefont {Choi}},
  \bibinfo {author} {\bibfnamefont {Q.}~\bibnamefont {Zhang}}, \bibinfo
  {author} {\bibfnamefont {W.}~\bibnamefont {Qin}}, \bibinfo {author}
  {\bibfnamefont {S.}~\bibnamefont {Yi}}, \bibinfo {author} {\bibfnamefont
  {P.}~\bibnamefont {Wang}}, \bibinfo {author} {\bibfnamefont {L.}~\bibnamefont
  {Li}}, \bibinfo {author} {\bibfnamefont {Y.}~\bibnamefont {Wang}}, \bibinfo
  {author} {\bibfnamefont {H.}~\bibnamefont {Zhang}}, \bibinfo {author}
  {\bibfnamefont {Z.}~\bibnamefont {Sun}}, \bibinfo {author} {\bibfnamefont
  {L.}~\bibnamefont {Wei}}, \bibinfo {author} {\bibfnamefont {S.}~\bibnamefont
  {Zhang}}, \bibinfo {author} {\bibfnamefont {T.}~\bibnamefont {Guo}}, \bibinfo
  {author} {\bibfnamefont {Q.}~\bibnamefont {Lu}}, \bibinfo {author}
  {\bibfnamefont {J.-H.}\ \bibnamefont {Cho}}, \bibinfo {author} {\bibfnamefont
  {C.}~\bibnamefont {Zeng}},\ and\ \bibinfo {author} {\bibfnamefont
  {Z.}~\bibnamefont {Zhang}},\ }\href@noop {} {\bibfield  {journal} {\bibinfo
  {journal} {Phys. Rev. Lett.}\ }\textbf {\bibinfo {volume} {121}},\ \bibinfo
  {pages} {096401} (\bibinfo {year} {2018})}\BibitemShut {NoStop}%
\bibitem [{\citenamefont {Yu}\ and\ \citenamefont {Li}(2012)}]{YuPRB2012}%
  \BibitemOpen
  \bibfield  {author} {\bibinfo {author} {\bibfnamefont {S.-L.}\ \bibnamefont
  {Yu}}\ and\ \bibinfo {author} {\bibfnamefont {J.-X.}\ \bibnamefont {Li}},\
  }\href@noop {} {\bibfield  {journal} {\bibinfo  {journal} {Phys. Rev. B}\
  }\textbf {\bibinfo {volume} {85}},\ \bibinfo {pages} {144402} (\bibinfo
  {year} {2012})}\BibitemShut {NoStop}%
\bibitem [{\citenamefont {Wang}\ \emph {et~al.}(2013)\citenamefont {Wang},
  \citenamefont {Li}, \citenamefont {Xiang},\ and\ \citenamefont
  {Wang}}]{WangPRB2013}%
  \BibitemOpen
  \bibfield  {author} {\bibinfo {author} {\bibfnamefont {W.-S.}\ \bibnamefont
  {Wang}}, \bibinfo {author} {\bibfnamefont {Z.-Z.}\ \bibnamefont {Li}},
  \bibinfo {author} {\bibfnamefont {Y.-Y.}\ \bibnamefont {Xiang}},\ and\
  \bibinfo {author} {\bibfnamefont {Q.-H.}\ \bibnamefont {Wang}},\ }\href
  {https://doi.org/10.1103/PhysRevB.87.115135} {\bibfield  {journal} {\bibinfo
  {journal} {Phys. Rev. B}\ }\textbf {\bibinfo {volume} {87}},\ \bibinfo
  {pages} {115135} (\bibinfo {year} {2013})}\BibitemShut {NoStop}%
\bibitem [{\citenamefont {Kiesel}\ and\ \citenamefont
  {Thomale}(2012)}]{KieselPRB2012}%
  \BibitemOpen
  \bibfield  {author} {\bibinfo {author} {\bibfnamefont {M.~L.}\ \bibnamefont
  {Kiesel}}\ and\ \bibinfo {author} {\bibfnamefont {R.}~\bibnamefont
  {Thomale}},\ }\href@noop {} {\bibfield  {journal} {\bibinfo  {journal} {Phys.
  Rev. B}\ }\textbf {\bibinfo {volume} {86}},\ \bibinfo {pages} {121105(R)}
  (\bibinfo {year} {2012})}\BibitemShut {NoStop}%
\bibitem [{\citenamefont {Kiesel}\ \emph {et~al.}(2013)\citenamefont {Kiesel},
  \citenamefont {Platt},\ and\ \citenamefont {Thomale}}]{KieselPRL2013}%
  \BibitemOpen
  \bibfield  {author} {\bibinfo {author} {\bibfnamefont {M.~L.}\ \bibnamefont
  {Kiesel}}, \bibinfo {author} {\bibfnamefont {C.}~\bibnamefont {Platt}},\ and\
  \bibinfo {author} {\bibfnamefont {R.}~\bibnamefont {Thomale}},\ }\href@noop
  {} {\bibfield  {journal} {\bibinfo  {journal} {Phys. Rev. Lett.}\ }\textbf
  {\bibinfo {volume} {110}},\ \bibinfo {pages} {126405} (\bibinfo {year}
  {2013})}\BibitemShut {NoStop}%
\bibitem [{\citenamefont {Ortiz}\ \emph {et~al.}(2019)\citenamefont {Ortiz},
  \citenamefont {Gomes}, \citenamefont {Morey}, \citenamefont {Winiarski},
  \citenamefont {Bordelon}, \citenamefont {Mangum}, \citenamefont {Oswald},
  \citenamefont {Rodriguez-Rivera}, \citenamefont {Neilson}, \citenamefont
  {Wilson}, \citenamefont {Ertekin}, \citenamefont {McQueen},\ and\
  \citenamefont {Toberer}}]{OrtizPRM2019}%
  \BibitemOpen
  \bibfield  {author} {\bibinfo {author} {\bibfnamefont {B.~R.}\ \bibnamefont
  {Ortiz}}, \bibinfo {author} {\bibfnamefont {L.~C.}\ \bibnamefont {Gomes}},
  \bibinfo {author} {\bibfnamefont {J.~R.}\ \bibnamefont {Morey}}, \bibinfo
  {author} {\bibfnamefont {M.}~\bibnamefont {Winiarski}}, \bibinfo {author}
  {\bibfnamefont {M.}~\bibnamefont {Bordelon}}, \bibinfo {author}
  {\bibfnamefont {J.~S.}\ \bibnamefont {Mangum}}, \bibinfo {author}
  {\bibfnamefont {I.~W.~H.}\ \bibnamefont {Oswald}}, \bibinfo {author}
  {\bibfnamefont {J.~A.}\ \bibnamefont {Rodriguez-Rivera}}, \bibinfo {author}
  {\bibfnamefont {J.~R.}\ \bibnamefont {Neilson}}, \bibinfo {author}
  {\bibfnamefont {S.~D.}\ \bibnamefont {Wilson}}, \bibinfo {author}
  {\bibfnamefont {E.}~\bibnamefont {Ertekin}}, \bibinfo {author} {\bibfnamefont
  {T.~M.}\ \bibnamefont {McQueen}},\ and\ \bibinfo {author} {\bibfnamefont
  {E.~S.}\ \bibnamefont {Toberer}},\ }\href@noop {} {\bibfield  {journal}
  {\bibinfo  {journal} {Phys. Rev. Mater.}\ }\textbf {\bibinfo {volume} {3}},\
  \bibinfo {pages} {094407} (\bibinfo {year} {2019})}\BibitemShut {NoStop}%
\bibitem [{\citenamefont {Ortiz}\ \emph
  {et~al.}(2021{\natexlab{a}})\citenamefont {Ortiz}, \citenamefont {Sarte},
  \citenamefont {Kenney}, \citenamefont {Graf}, \citenamefont {Teicher},
  \citenamefont {Seshadri},\ and\ \citenamefont {Wilson}}]{OrtizPRM2021}%
  \BibitemOpen
  \bibfield  {author} {\bibinfo {author} {\bibfnamefont {B.~R.}\ \bibnamefont
  {Ortiz}}, \bibinfo {author} {\bibfnamefont {P.~M.}\ \bibnamefont {Sarte}},
  \bibinfo {author} {\bibfnamefont {E.~M.}\ \bibnamefont {Kenney}}, \bibinfo
  {author} {\bibfnamefont {M.~J.}\ \bibnamefont {Graf}}, \bibinfo {author}
  {\bibfnamefont {S.~M.~L.}\ \bibnamefont {Teicher}}, \bibinfo {author}
  {\bibfnamefont {R.}~\bibnamefont {Seshadri}},\ and\ \bibinfo {author}
  {\bibfnamefont {S.~D.}\ \bibnamefont {Wilson}},\ }\href@noop {} {\bibfield
  {journal} {\bibinfo  {journal} {Phys. Rev. Mater.}\ }\textbf {\bibinfo
  {volume} {5}},\ \bibinfo {pages} {034801} (\bibinfo {year}
  {2021}{\natexlab{a}})}\BibitemShut {NoStop}%
\bibitem [{\citenamefont {Ortiz}\ \emph {et~al.}(2020)\citenamefont {Ortiz},
  \citenamefont {Teicher}, \citenamefont {Hu}, \citenamefont {Zuo},
  \citenamefont {Sarte}, \citenamefont {Schueller}, \citenamefont {Abeykoon},
  \citenamefont {Krogstad}, \citenamefont {Rosenkranz}, \citenamefont {Osborn},
  \citenamefont {Seshadri}, \citenamefont {Balents}, \citenamefont {He},\ and\
  \citenamefont {Wilson}}]{OrtizPRL2020}%
  \BibitemOpen
  \bibfield  {author} {\bibinfo {author} {\bibfnamefont {B.~R.}\ \bibnamefont
  {Ortiz}}, \bibinfo {author} {\bibfnamefont {S.~M.~L.}\ \bibnamefont
  {Teicher}}, \bibinfo {author} {\bibfnamefont {Y.}~\bibnamefont {Hu}},
  \bibinfo {author} {\bibfnamefont {J.~L.}\ \bibnamefont {Zuo}}, \bibinfo
  {author} {\bibfnamefont {P.~M.}\ \bibnamefont {Sarte}}, \bibinfo {author}
  {\bibfnamefont {E.~C.}\ \bibnamefont {Schueller}}, \bibinfo {author}
  {\bibfnamefont {A.~M.~M.}\ \bibnamefont {Abeykoon}}, \bibinfo {author}
  {\bibfnamefont {M.~J.}\ \bibnamefont {Krogstad}}, \bibinfo {author}
  {\bibfnamefont {S.}~\bibnamefont {Rosenkranz}}, \bibinfo {author}
  {\bibfnamefont {R.}~\bibnamefont {Osborn}}, \bibinfo {author} {\bibfnamefont
  {R.}~\bibnamefont {Seshadri}}, \bibinfo {author} {\bibfnamefont
  {L.}~\bibnamefont {Balents}}, \bibinfo {author} {\bibfnamefont
  {J.}~\bibnamefont {He}},\ and\ \bibinfo {author} {\bibfnamefont {S.~D.}\
  \bibnamefont {Wilson}},\ }\href@noop {} {\bibfield  {journal} {\bibinfo
  {journal} {Phys. Rev. Lett.}\ }\textbf {\bibinfo {volume} {125}},\ \bibinfo
  {pages} {247002} (\bibinfo {year} {2020})}\BibitemShut {NoStop}%
\bibitem [{\citenamefont {Yin}\ \emph {et~al.}(2021)\citenamefont {Yin},
  \citenamefont {Tu}, \citenamefont {Gong}, \citenamefont {Fu}, \citenamefont
  {Yan},\ and\ \citenamefont {Lei}}]{YinCPL2021}%
  \BibitemOpen
  \bibfield  {author} {\bibinfo {author} {\bibfnamefont {Q.}~\bibnamefont
  {Yin}}, \bibinfo {author} {\bibfnamefont {Z.}~\bibnamefont {Tu}}, \bibinfo
  {author} {\bibfnamefont {C.}~\bibnamefont {Gong}}, \bibinfo {author}
  {\bibfnamefont {Y.}~\bibnamefont {Fu}}, \bibinfo {author} {\bibfnamefont
  {S.}~\bibnamefont {Yan}},\ and\ \bibinfo {author} {\bibfnamefont
  {H.}~\bibnamefont {Lei}},\ }\href@noop {} {\bibfield  {journal} {\bibinfo
  {journal} {Chin. Phys. Lett.}\ }\textbf {\bibinfo {volume} {38}},\ \bibinfo
  {pages} {037403} (\bibinfo {year} {2021})}\BibitemShut {NoStop}%
\bibitem [{\citenamefont {Liang}\ \emph {et~al.}(2021)\citenamefont {Liang},
  \citenamefont {Hou}, \citenamefont {Zhang}, \citenamefont {Ma}, \citenamefont
  {Wu}, \citenamefont {Zhang}, \citenamefont {Yu}, \citenamefont {Ying},
  \citenamefont {Jiang}, \citenamefont {Shan}, \citenamefont {Wang},\ and\
  \citenamefont {Chen}}]{LiangPRX2021}%
  \BibitemOpen
  \bibfield  {author} {\bibinfo {author} {\bibfnamefont {Z.}~\bibnamefont
  {Liang}}, \bibinfo {author} {\bibfnamefont {X.}~\bibnamefont {Hou}}, \bibinfo
  {author} {\bibfnamefont {F.}~\bibnamefont {Zhang}}, \bibinfo {author}
  {\bibfnamefont {W.}~\bibnamefont {Ma}}, \bibinfo {author} {\bibfnamefont
  {P.}~\bibnamefont {Wu}}, \bibinfo {author} {\bibfnamefont {Z.}~\bibnamefont
  {Zhang}}, \bibinfo {author} {\bibfnamefont {F.}~\bibnamefont {Yu}}, \bibinfo
  {author} {\bibfnamefont {J.-J.}\ \bibnamefont {Ying}}, \bibinfo {author}
  {\bibfnamefont {K.}~\bibnamefont {Jiang}}, \bibinfo {author} {\bibfnamefont
  {L.}~\bibnamefont {Shan}}, \bibinfo {author} {\bibfnamefont {Z.}~\bibnamefont
  {Wang}},\ and\ \bibinfo {author} {\bibfnamefont {X.-H.}\ \bibnamefont
  {Chen}},\ }\href@noop {} {\bibfield  {journal} {\bibinfo  {journal} {Phys.
  Rev. X}\ }\textbf {\bibinfo {volume} {11}},\ \bibinfo {pages} {031026}
  (\bibinfo {year} {2021})}\BibitemShut {NoStop}%
\bibitem [{\citenamefont {Li}\ \emph {et~al.}(2021)\citenamefont {Li},
  \citenamefont {Zhang}, \citenamefont {Yilmaz}, \citenamefont {Pai},
  \citenamefont {Marvinney}, \citenamefont {Said}, \citenamefont {Yin},
  \citenamefont {Gong}, \citenamefont {Tu}, \citenamefont {Vescovo},
  \citenamefont {Nelson}, \citenamefont {Moore}, \citenamefont {Murakami},
  \citenamefont {Lei}, \citenamefont {Lee}, \citenamefont {Lawrie},\ and\
  \citenamefont {Miao}}]{LiPRX2021}%
  \BibitemOpen
  \bibfield  {author} {\bibinfo {author} {\bibfnamefont {H.}~\bibnamefont
  {Li}}, \bibinfo {author} {\bibfnamefont {T.~T.}\ \bibnamefont {Zhang}},
  \bibinfo {author} {\bibfnamefont {T.}~\bibnamefont {Yilmaz}}, \bibinfo
  {author} {\bibfnamefont {Y.~Y.}\ \bibnamefont {Pai}}, \bibinfo {author}
  {\bibfnamefont {C.~E.}\ \bibnamefont {Marvinney}}, \bibinfo {author}
  {\bibfnamefont {A.}~\bibnamefont {Said}}, \bibinfo {author} {\bibfnamefont
  {Q.~W.}\ \bibnamefont {Yin}}, \bibinfo {author} {\bibfnamefont {C.~S.}\
  \bibnamefont {Gong}}, \bibinfo {author} {\bibfnamefont {Z.~J.}\ \bibnamefont
  {Tu}}, \bibinfo {author} {\bibfnamefont {E.}~\bibnamefont {Vescovo}},
  \bibinfo {author} {\bibfnamefont {C.~S.}\ \bibnamefont {Nelson}}, \bibinfo
  {author} {\bibfnamefont {R.~G.}\ \bibnamefont {Moore}}, \bibinfo {author}
  {\bibfnamefont {S.}~\bibnamefont {Murakami}}, \bibinfo {author}
  {\bibfnamefont {H.~C.}\ \bibnamefont {Lei}}, \bibinfo {author} {\bibfnamefont
  {H.~N.}\ \bibnamefont {Lee}}, \bibinfo {author} {\bibfnamefont {B.~J.}\
  \bibnamefont {Lawrie}},\ and\ \bibinfo {author} {\bibfnamefont
  {H.}~\bibnamefont {Miao}},\ }\href@noop {} {\bibfield  {journal} {\bibinfo
  {journal} {Phys. Rev. X}\ }\textbf {\bibinfo {volume} {11}},\ \bibinfo
  {pages} {031050} (\bibinfo {year} {2021})}\BibitemShut {NoStop}%
\bibitem [{\citenamefont {Ortiz}\ \emph
  {et~al.}(2021{\natexlab{b}})\citenamefont {Ortiz}, \citenamefont {Teicher},
  \citenamefont {Kautzsch}, \citenamefont {Sarte}, \citenamefont {Ratcliff},
  \citenamefont {Harter}, \citenamefont {Ruff}, \citenamefont {Seshadri},\ and\
  \citenamefont {Wilson}}]{OrtizPRX2021}%
  \BibitemOpen
  \bibfield  {author} {\bibinfo {author} {\bibfnamefont {B.~R.}\ \bibnamefont
  {Ortiz}}, \bibinfo {author} {\bibfnamefont {S.~M.~L.}\ \bibnamefont
  {Teicher}}, \bibinfo {author} {\bibfnamefont {L.}~\bibnamefont {Kautzsch}},
  \bibinfo {author} {\bibfnamefont {P.~M.}\ \bibnamefont {Sarte}}, \bibinfo
  {author} {\bibfnamefont {N.}~\bibnamefont {Ratcliff}}, \bibinfo {author}
  {\bibfnamefont {J.}~\bibnamefont {Harter}}, \bibinfo {author} {\bibfnamefont
  {J.~P.~C.}\ \bibnamefont {Ruff}}, \bibinfo {author} {\bibfnamefont
  {R.}~\bibnamefont {Seshadri}},\ and\ \bibinfo {author} {\bibfnamefont
  {S.~D.}\ \bibnamefont {Wilson}},\ }\href@noop {} {\bibfield  {journal}
  {\bibinfo  {journal} {Phys. Rev. X}\ }\textbf {\bibinfo {volume} {11}},\
  \bibinfo {pages} {041030} (\bibinfo {year} {2021}{\natexlab{b}})}\BibitemShut
  {NoStop}%
\bibitem [{\citenamefont {Du}\ \emph {et~al.}(2021)\citenamefont {Du},
  \citenamefont {Luo}, \citenamefont {Ortiz}, \citenamefont {Chen},
  \citenamefont {Duan}, \citenamefont {Zhang}, \citenamefont {Lu},
  \citenamefont {Wilson}, \citenamefont {Song},\ and\ \citenamefont
  {Yuan}}]{DuPRB2021}%
  \BibitemOpen
  \bibfield  {author} {\bibinfo {author} {\bibfnamefont {F.}~\bibnamefont
  {Du}}, \bibinfo {author} {\bibfnamefont {S.}~\bibnamefont {Luo}}, \bibinfo
  {author} {\bibfnamefont {B.~R.}\ \bibnamefont {Ortiz}}, \bibinfo {author}
  {\bibfnamefont {Y.}~\bibnamefont {Chen}}, \bibinfo {author} {\bibfnamefont
  {W.}~\bibnamefont {Duan}}, \bibinfo {author} {\bibfnamefont {D.}~\bibnamefont
  {Zhang}}, \bibinfo {author} {\bibfnamefont {X.}~\bibnamefont {Lu}}, \bibinfo
  {author} {\bibfnamefont {S.~D.}\ \bibnamefont {Wilson}}, \bibinfo {author}
  {\bibfnamefont {Y.}~\bibnamefont {Song}},\ and\ \bibinfo {author}
  {\bibfnamefont {H.}~\bibnamefont {Yuan}},\ }\href@noop {} {\bibfield
  {journal} {\bibinfo  {journal} {Phys. Rev. B}\ }\textbf {\bibinfo {volume}
  {103}},\ \bibinfo {pages} {L220504} (\bibinfo {year} {2021})}\BibitemShut
  {NoStop}%
\bibitem [{\citenamefont {Chen}\ \emph
  {et~al.}(2021{\natexlab{a}})\citenamefont {Chen}, \citenamefont {Wang},
  \citenamefont {Yin}, \citenamefont {Gu}, \citenamefont {Jiang}, \citenamefont
  {Tu}, \citenamefont {Gong}, \citenamefont {Uwatoko}, \citenamefont {Sun},
  \citenamefont {Lei}, \citenamefont {Hu},\ and\ \citenamefont
  {Cheng}}]{ChenPRL2021}%
  \BibitemOpen
  \bibfield  {author} {\bibinfo {author} {\bibfnamefont {K.~Y.}\ \bibnamefont
  {Chen}}, \bibinfo {author} {\bibfnamefont {N.~N.}\ \bibnamefont {Wang}},
  \bibinfo {author} {\bibfnamefont {Q.~W.}\ \bibnamefont {Yin}}, \bibinfo
  {author} {\bibfnamefont {Y.~H.}\ \bibnamefont {Gu}}, \bibinfo {author}
  {\bibfnamefont {K.}~\bibnamefont {Jiang}}, \bibinfo {author} {\bibfnamefont
  {Z.~J.}\ \bibnamefont {Tu}}, \bibinfo {author} {\bibfnamefont {C.~S.}\
  \bibnamefont {Gong}}, \bibinfo {author} {\bibfnamefont {Y.}~\bibnamefont
  {Uwatoko}}, \bibinfo {author} {\bibfnamefont {J.~P.}\ \bibnamefont {Sun}},
  \bibinfo {author} {\bibfnamefont {H.~C.}\ \bibnamefont {Lei}}, \bibinfo
  {author} {\bibfnamefont {J.~P.}\ \bibnamefont {Hu}},\ and\ \bibinfo {author}
  {\bibfnamefont {J.-G.}\ \bibnamefont {Cheng}},\ }\href@noop {} {\bibfield
  {journal} {\bibinfo  {journal} {Phys. Rev. Lett.}\ }\textbf {\bibinfo
  {volume} {126}},\ \bibinfo {pages} {247001} (\bibinfo {year}
  {2021}{\natexlab{a}})}\BibitemShut {NoStop}%
\bibitem [{\citenamefont {Zhang}\ \emph {et~al.}(2021)\citenamefont {Zhang},
  \citenamefont {Chen}, \citenamefont {Zhou}, \citenamefont {Yuan},
  \citenamefont {Wang}, \citenamefont {Wang}, \citenamefont {Yang},
  \citenamefont {An}, \citenamefont {Zhang}, \citenamefont {Zhu}, \citenamefont
  {Zhou}, \citenamefont {Chen}, \citenamefont {Zhou},\ and\ \citenamefont
  {Yang}}]{ZhangPRB2021}%
  \BibitemOpen
  \bibfield  {author} {\bibinfo {author} {\bibfnamefont {Z.}~\bibnamefont
  {Zhang}}, \bibinfo {author} {\bibfnamefont {Z.}~\bibnamefont {Chen}},
  \bibinfo {author} {\bibfnamefont {Y.}~\bibnamefont {Zhou}}, \bibinfo {author}
  {\bibfnamefont {Y.}~\bibnamefont {Yuan}}, \bibinfo {author} {\bibfnamefont
  {S.}~\bibnamefont {Wang}}, \bibinfo {author} {\bibfnamefont {J.}~\bibnamefont
  {Wang}}, \bibinfo {author} {\bibfnamefont {H.}~\bibnamefont {Yang}}, \bibinfo
  {author} {\bibfnamefont {C.}~\bibnamefont {An}}, \bibinfo {author}
  {\bibfnamefont {L.}~\bibnamefont {Zhang}}, \bibinfo {author} {\bibfnamefont
  {X.}~\bibnamefont {Zhu}}, \bibinfo {author} {\bibfnamefont {Y.}~\bibnamefont
  {Zhou}}, \bibinfo {author} {\bibfnamefont {X.}~\bibnamefont {Chen}}, \bibinfo
  {author} {\bibfnamefont {J.}~\bibnamefont {Zhou}},\ and\ \bibinfo {author}
  {\bibfnamefont {Z.}~\bibnamefont {Yang}},\ }\href@noop {} {\bibfield
  {journal} {\bibinfo  {journal} {Phys. Rev. B}\ }\textbf {\bibinfo {volume}
  {103}},\ \bibinfo {pages} {224513} (\bibinfo {year} {2021})}\BibitemShut
  {NoStop}%
\bibitem [{\citenamefont {Du}\ \emph {et~al.}(2022)\citenamefont {Du},
  \citenamefont {Luo}, \citenamefont {Li}, \citenamefont {Ortiz}, \citenamefont
  {Chen}, \citenamefont {Wilson}, \citenamefont {Song},\ and\ \citenamefont
  {Yuan}}]{DuCPB2021}%
  \BibitemOpen
  \bibfield  {author} {\bibinfo {author} {\bibfnamefont {F.}~\bibnamefont
  {Du}}, \bibinfo {author} {\bibfnamefont {S.}~\bibnamefont {Luo}}, \bibinfo
  {author} {\bibfnamefont {R.}~\bibnamefont {Li}}, \bibinfo {author}
  {\bibfnamefont {B.~R.}\ \bibnamefont {Ortiz}}, \bibinfo {author}
  {\bibfnamefont {Y.}~\bibnamefont {Chen}}, \bibinfo {author} {\bibfnamefont
  {S.~D.}\ \bibnamefont {Wilson}}, \bibinfo {author} {\bibfnamefont
  {Y.}~\bibnamefont {Song}},\ and\ \bibinfo {author} {\bibfnamefont
  {H.}~\bibnamefont {Yuan}},\ }\href@noop {} {\bibfield  {journal} {\bibinfo
  {journal} {Chin. Phys. B}\ }\textbf {\bibinfo {volume} {31}},\ \bibinfo
  {pages} {017404} (\bibinfo {year} {2022})}\BibitemShut {NoStop}%
\bibitem [{\citenamefont {Nakayama}\ \emph {et~al.}(2022)\citenamefont
  {Nakayama}, \citenamefont {Li}, \citenamefont {Kato}, \citenamefont {Liu},
  \citenamefont {Wang}, \citenamefont {Takahashi}, \citenamefont {Yao},\ and\
  \citenamefont {Sato}}]{NakayamaPRX2022}%
  \BibitemOpen
  \bibfield  {author} {\bibinfo {author} {\bibfnamefont {K.}~\bibnamefont
  {Nakayama}}, \bibinfo {author} {\bibfnamefont {Y.}~\bibnamefont {Li}},
  \bibinfo {author} {\bibfnamefont {T.}~\bibnamefont {Kato}}, \bibinfo {author}
  {\bibfnamefont {M.}~\bibnamefont {Liu}}, \bibinfo {author} {\bibfnamefont
  {Z.}~\bibnamefont {Wang}}, \bibinfo {author} {\bibfnamefont {T.}~\bibnamefont
  {Takahashi}}, \bibinfo {author} {\bibfnamefont {Y.}~\bibnamefont {Yao}},\
  and\ \bibinfo {author} {\bibfnamefont {T.}~\bibnamefont {Sato}},\ }\href@noop
  {} {\bibfield  {journal} {\bibinfo  {journal} {Phys. Rev. X}\ }\textbf
  {\bibinfo {volume} {12}},\ \bibinfo {pages} {011001} (\bibinfo {year}
  {2022})}\BibitemShut {NoStop}%
\bibitem [{\citenamefont {Song}\ \emph {et~al.}(2021)\citenamefont {Song},
  \citenamefont {Ying}, \citenamefont {Chen}, \citenamefont {Han},
  \citenamefont {Wu}, \citenamefont {Schnyder}, \citenamefont {Huang},
  \citenamefont {g.~Guo},\ and\ \citenamefont {Chen}}]{SongPRL2021}%
  \BibitemOpen
  \bibfield  {author} {\bibinfo {author} {\bibfnamefont {Y.}~\bibnamefont
  {Song}}, \bibinfo {author} {\bibfnamefont {T.}~\bibnamefont {Ying}}, \bibinfo
  {author} {\bibfnamefont {X.}~\bibnamefont {Chen}}, \bibinfo {author}
  {\bibfnamefont {X.}~\bibnamefont {Han}}, \bibinfo {author} {\bibfnamefont
  {X.}~\bibnamefont {Wu}}, \bibinfo {author} {\bibfnamefont {A.~P.}\
  \bibnamefont {Schnyder}}, \bibinfo {author} {\bibfnamefont {Y.}~\bibnamefont
  {Huang}}, \bibinfo {author} {\bibfnamefont {J.}~\bibnamefont {g.~Guo}},\ and\
  \bibinfo {author} {\bibfnamefont {X.}~\bibnamefont {Chen}},\ }\href@noop {}
  {\bibfield  {journal} {\bibinfo  {journal} {Phys. Rev. Lett.}\ }\textbf
  {\bibinfo {volume} {127}},\ \bibinfo {pages} {237001} (\bibinfo {year}
  {2021})}\BibitemShut {NoStop}%
\bibitem [{\citenamefont {Oey}\ \emph {et~al.}(2022{\natexlab{a}})\citenamefont
  {Oey}, \citenamefont {Ortiz}, \citenamefont {Kaboudvand}, \citenamefont
  {Frassineti}, \citenamefont {Garcia}, \citenamefont {Cong}, \citenamefont
  {Sanna}, \citenamefont {Mitrovi\ifmmode~\acute{c}\else \'{c}\fi{}},
  \citenamefont {Seshadri},\ and\ \citenamefont {Wilson}}]{OeyPRM2022}%
  \BibitemOpen
  \bibfield  {author} {\bibinfo {author} {\bibfnamefont {Y.~M.}\ \bibnamefont
  {Oey}}, \bibinfo {author} {\bibfnamefont {B.~R.}\ \bibnamefont {Ortiz}},
  \bibinfo {author} {\bibfnamefont {F.}~\bibnamefont {Kaboudvand}}, \bibinfo
  {author} {\bibfnamefont {J.}~\bibnamefont {Frassineti}}, \bibinfo {author}
  {\bibfnamefont {E.}~\bibnamefont {Garcia}}, \bibinfo {author} {\bibfnamefont
  {R.}~\bibnamefont {Cong}}, \bibinfo {author} {\bibfnamefont {S.}~\bibnamefont
  {Sanna}}, \bibinfo {author} {\bibfnamefont {V.~F.}\ \bibnamefont
  {Mitrovi\ifmmode~\acute{c}\else \'{c}\fi{}}}, \bibinfo {author}
  {\bibfnamefont {R.}~\bibnamefont {Seshadri}},\ and\ \bibinfo {author}
  {\bibfnamefont {S.~D.}\ \bibnamefont {Wilson}},\ }\href
  {https://doi.org/10.1103/PhysRevMaterials.6.L041801} {\bibfield  {journal}
  {\bibinfo  {journal} {Phys. Rev. Mater.}\ }\textbf {\bibinfo {volume} {6}},\
  \bibinfo {pages} {L041801} (\bibinfo {year}
  {2022}{\natexlab{a}})}\BibitemShut {NoStop}%
\bibitem [{\citenamefont {Oey}\ \emph {et~al.}(2022{\natexlab{b}})\citenamefont
  {Oey}, \citenamefont {Kaboudvand}, \citenamefont {Ortiz}, \citenamefont
  {Seshadri},\ and\ \citenamefont {Wilson}}]{OeyPRM2022_2}%
  \BibitemOpen
  \bibfield  {author} {\bibinfo {author} {\bibfnamefont {Y.~M.}\ \bibnamefont
  {Oey}}, \bibinfo {author} {\bibfnamefont {F.}~\bibnamefont {Kaboudvand}},
  \bibinfo {author} {\bibfnamefont {B.~R.}\ \bibnamefont {Ortiz}}, \bibinfo
  {author} {\bibfnamefont {R.}~\bibnamefont {Seshadri}},\ and\ \bibinfo
  {author} {\bibfnamefont {S.~D.}\ \bibnamefont {Wilson}},\ }\href
  {https://doi.org/10.1103/PhysRevMaterials.6.074802} {\bibfield  {journal}
  {\bibinfo  {journal} {Phys. Rev. Mater.}\ }\textbf {\bibinfo {volume} {6}},\
  \bibinfo {pages} {074802} (\bibinfo {year} {2022}{\natexlab{b}})}\BibitemShut
  {NoStop}%
\bibitem [{\citenamefont {Liu}\ \emph {et~al.}()\citenamefont {Liu},
  \citenamefont {Wang}, \citenamefont {Cai}, \citenamefont {Hao}, \citenamefont
  {Ma}, \citenamefont {Wang}, \citenamefont {Liu}, \citenamefont {Chen},
  \citenamefont {Zhou}, \citenamefont {Wang}, \citenamefont {Wang},
  \citenamefont {He}, \citenamefont {Liu}, \citenamefont {Cui}, \citenamefont
  {Wang}, \citenamefont {Huang}, \citenamefont {Chen},\ and\ \citenamefont
  {Mei}}]{LiuarXiv2021}%
  \BibitemOpen
  \bibfield  {author} {\bibinfo {author} {\bibfnamefont {Y.}~\bibnamefont
  {Liu}}, \bibinfo {author} {\bibfnamefont {Y.}~\bibnamefont {Wang}}, \bibinfo
  {author} {\bibfnamefont {Y.}~\bibnamefont {Cai}}, \bibinfo {author}
  {\bibfnamefont {Z.}~\bibnamefont {Hao}}, \bibinfo {author} {\bibfnamefont
  {X.-M.}\ \bibnamefont {Ma}}, \bibinfo {author} {\bibfnamefont
  {L.}~\bibnamefont {Wang}}, \bibinfo {author} {\bibfnamefont {C.}~\bibnamefont
  {Liu}}, \bibinfo {author} {\bibfnamefont {J.}~\bibnamefont {Chen}}, \bibinfo
  {author} {\bibfnamefont {L.}~\bibnamefont {Zhou}}, \bibinfo {author}
  {\bibfnamefont {J.}~\bibnamefont {Wang}}, \bibinfo {author} {\bibfnamefont
  {S.}~\bibnamefont {Wang}}, \bibinfo {author} {\bibfnamefont {H.}~\bibnamefont
  {He}}, \bibinfo {author} {\bibfnamefont {Y.}~\bibnamefont {Liu}}, \bibinfo
  {author} {\bibfnamefont {S.}~\bibnamefont {Cui}}, \bibinfo {author}
  {\bibfnamefont {J.}~\bibnamefont {Wang}}, \bibinfo {author} {\bibfnamefont
  {B.}~\bibnamefont {Huang}}, \bibinfo {author} {\bibfnamefont
  {C.}~\bibnamefont {Chen}},\ and\ \bibinfo {author} {\bibfnamefont {J.-W.}\
  \bibnamefont {Mei}},\ }\bibinfo {note} {arXiv:2110.12651}\BibitemShut
  {NoStop}%
\bibitem [{\citenamefont {Yang}\ \emph {et~al.}(2022)\citenamefont {Yang},
  \citenamefont {Huang}, \citenamefont {Zhang}, \citenamefont {Zhao},
  \citenamefont {Shi}, \citenamefont {Luo}, \citenamefont {Zhao}, \citenamefont
  {Qian}, \citenamefont {Tan}, \citenamefont {Hu}, \citenamefont {Zhu},
  \citenamefont {Lu}, \citenamefont {Zhang}, \citenamefont {Sun}, \citenamefont
  {Cheng}, \citenamefont {Shen}, \citenamefont {Lin}, \citenamefont {Yan},
  \citenamefont {Zhou}, \citenamefont {Wang}, \citenamefont {Pennycook},
  \citenamefont {Chen}, \citenamefont {Dong}, \citenamefont {Zhou},\ and\
  \citenamefont {Gao}}]{YangSB2022}%
  \BibitemOpen
  \bibfield  {author} {\bibinfo {author} {\bibfnamefont {H.}~\bibnamefont
  {Yang}}, \bibinfo {author} {\bibfnamefont {Z.}~\bibnamefont {Huang}},
  \bibinfo {author} {\bibfnamefont {Y.}~\bibnamefont {Zhang}}, \bibinfo
  {author} {\bibfnamefont {Z.}~\bibnamefont {Zhao}}, \bibinfo {author}
  {\bibfnamefont {J.}~\bibnamefont {Shi}}, \bibinfo {author} {\bibfnamefont
  {H.}~\bibnamefont {Luo}}, \bibinfo {author} {\bibfnamefont {L.}~\bibnamefont
  {Zhao}}, \bibinfo {author} {\bibfnamefont {G.}~\bibnamefont {Qian}}, \bibinfo
  {author} {\bibfnamefont {H.}~\bibnamefont {Tan}}, \bibinfo {author}
  {\bibfnamefont {B.}~\bibnamefont {Hu}}, \bibinfo {author} {\bibfnamefont
  {K.}~\bibnamefont {Zhu}}, \bibinfo {author} {\bibfnamefont {Z.}~\bibnamefont
  {Lu}}, \bibinfo {author} {\bibfnamefont {H.}~\bibnamefont {Zhang}}, \bibinfo
  {author} {\bibfnamefont {J.}~\bibnamefont {Sun}}, \bibinfo {author}
  {\bibfnamefont {J.}~\bibnamefont {Cheng}}, \bibinfo {author} {\bibfnamefont
  {C.}~\bibnamefont {Shen}}, \bibinfo {author} {\bibfnamefont {X.}~\bibnamefont
  {Lin}}, \bibinfo {author} {\bibfnamefont {B.}~\bibnamefont {Yan}}, \bibinfo
  {author} {\bibfnamefont {X.}~\bibnamefont {Zhou}}, \bibinfo {author}
  {\bibfnamefont {Z.}~\bibnamefont {Wang}}, \bibinfo {author} {\bibfnamefont
  {S.~J.}\ \bibnamefont {Pennycook}}, \bibinfo {author} {\bibfnamefont
  {H.}~\bibnamefont {Chen}}, \bibinfo {author} {\bibfnamefont {X.}~\bibnamefont
  {Dong}}, \bibinfo {author} {\bibfnamefont {W.}~\bibnamefont {Zhou}},\ and\
  \bibinfo {author} {\bibfnamefont {H.-J.}\ \bibnamefont {Gao}},\ }\href@noop
  {} {\bibfield  {journal} {\bibinfo  {journal} {Sci. Bull.}\ }\textbf
  {\bibinfo {volume} {67}},\ \bibinfo {pages} {2176} (\bibinfo {year}
  {2022})}\BibitemShut {NoStop}%
\bibitem [{\citenamefont {Qian}\ \emph {et~al.}(2021)\citenamefont {Qian},
  \citenamefont {Christensen}, \citenamefont {Hu}, \citenamefont {Saha},
  \citenamefont {Andersen}, \citenamefont {Fernandes}, \citenamefont {Birol},\
  and\ \citenamefont {Ni}}]{QianPRB2021}%
  \BibitemOpen
  \bibfield  {author} {\bibinfo {author} {\bibfnamefont {T.}~\bibnamefont
  {Qian}}, \bibinfo {author} {\bibfnamefont {M.~H.}\ \bibnamefont
  {Christensen}}, \bibinfo {author} {\bibfnamefont {C.}~\bibnamefont {Hu}},
  \bibinfo {author} {\bibfnamefont {A.}~\bibnamefont {Saha}}, \bibinfo {author}
  {\bibfnamefont {B.~M.}\ \bibnamefont {Andersen}}, \bibinfo {author}
  {\bibfnamefont {R.~M.}\ \bibnamefont {Fernandes}}, \bibinfo {author}
  {\bibfnamefont {T.}~\bibnamefont {Birol}},\ and\ \bibinfo {author}
  {\bibfnamefont {N.}~\bibnamefont {Ni}},\ }\href@noop {} {\bibfield  {journal}
  {\bibinfo  {journal} {Phys. Rev. B}\ }\textbf {\bibinfo {volume} {104}},\
  \bibinfo {pages} {144506} (\bibinfo {year} {2021})}\BibitemShut {NoStop}%
\bibitem [{\citenamefont {Kato}\ \emph
  {et~al.}(2022{\natexlab{a}})\citenamefont {Kato}, \citenamefont {Li},
  \citenamefont {Nakayama}, \citenamefont {Wang}, \citenamefont {Souma},
  \citenamefont {Matsui}, \citenamefont {Kitamura}, \citenamefont {Horiba},
  \citenamefont {Kumigashira}, \citenamefont {Takahashi}, \citenamefont {Yao},\
  and\ \citenamefont {Sato}}]{KatoPRL2022}%
  \BibitemOpen
  \bibfield  {author} {\bibinfo {author} {\bibfnamefont {T.}~\bibnamefont
  {Kato}}, \bibinfo {author} {\bibfnamefont {Y.}~\bibnamefont {Li}}, \bibinfo
  {author} {\bibfnamefont {K.}~\bibnamefont {Nakayama}}, \bibinfo {author}
  {\bibfnamefont {Z.}~\bibnamefont {Wang}}, \bibinfo {author} {\bibfnamefont
  {S.}~\bibnamefont {Souma}}, \bibinfo {author} {\bibfnamefont
  {F.}~\bibnamefont {Matsui}}, \bibinfo {author} {\bibfnamefont
  {M.}~\bibnamefont {Kitamura}}, \bibinfo {author} {\bibfnamefont
  {K.}~\bibnamefont {Horiba}}, \bibinfo {author} {\bibfnamefont
  {H.}~\bibnamefont {Kumigashira}}, \bibinfo {author} {\bibfnamefont
  {T.}~\bibnamefont {Takahashi}}, \bibinfo {author} {\bibfnamefont
  {Y.}~\bibnamefont {Yao}},\ and\ \bibinfo {author} {\bibfnamefont
  {T.}~\bibnamefont {Sato}},\ }\href
  {https://doi.org/10.1103/PhysRevLett.129.206402} {\bibfield  {journal}
  {\bibinfo  {journal} {Phys. Rev. Lett.}\ }\textbf {\bibinfo {volume} {129}},\
  \bibinfo {pages} {206402} (\bibinfo {year} {2022}{\natexlab{a}})}\BibitemShut
  {NoStop}%
\bibitem [{\citenamefont {Liu}\ \emph {et~al.}(2022)\citenamefont {Liu},
  \citenamefont {Han}, \citenamefont {Hu}, \citenamefont {Tu}, \citenamefont
  {Zhang}, \citenamefont {Long}, \citenamefont {Hou}, \citenamefont {Mu},\ and\
  \citenamefont {Shan}}]{LiuPRB2022}%
  \BibitemOpen
  \bibfield  {author} {\bibinfo {author} {\bibfnamefont {M.}~\bibnamefont
  {Liu}}, \bibinfo {author} {\bibfnamefont {T.}~\bibnamefont {Han}}, \bibinfo
  {author} {\bibfnamefont {X.}~\bibnamefont {Hu}}, \bibinfo {author}
  {\bibfnamefont {Y.}~\bibnamefont {Tu}}, \bibinfo {author} {\bibfnamefont
  {Z.}~\bibnamefont {Zhang}}, \bibinfo {author} {\bibfnamefont
  {M.}~\bibnamefont {Long}}, \bibinfo {author} {\bibfnamefont {X.}~\bibnamefont
  {Hou}}, \bibinfo {author} {\bibfnamefont {Q.}~\bibnamefont {Mu}},\ and\
  \bibinfo {author} {\bibfnamefont {L.}~\bibnamefont {Shan}},\ }\href
  {https://doi.org/10.1103/PhysRevB.106.L140501} {\bibfield  {journal}
  {\bibinfo  {journal} {Phys. Rev. B}\ }\textbf {\bibinfo {volume} {106}},\
  \bibinfo {pages} {L140501} (\bibinfo {year} {2022})}\BibitemShut {NoStop}%
\bibitem [{\citenamefont {Li}\ \emph {et~al.}(2022{\natexlab{a}})\citenamefont
  {Li}, \citenamefont {Li}, \citenamefont {Fan}, \citenamefont {Liu},
  \citenamefont {Feng}, \citenamefont {Liu}, \citenamefont {Wang},
  \citenamefont {Yin}, \citenamefont {Duan}, \citenamefont {Li}, \citenamefont
  {Wang}, \citenamefont {Wen},\ and\ \citenamefont {Yao}}]{LiPRB2022}%
  \BibitemOpen
  \bibfield  {author} {\bibinfo {author} {\bibfnamefont {Y.}~\bibnamefont
  {Li}}, \bibinfo {author} {\bibfnamefont {Q.}~\bibnamefont {Li}}, \bibinfo
  {author} {\bibfnamefont {X.}~\bibnamefont {Fan}}, \bibinfo {author}
  {\bibfnamefont {J.}~\bibnamefont {Liu}}, \bibinfo {author} {\bibfnamefont
  {Q.}~\bibnamefont {Feng}}, \bibinfo {author} {\bibfnamefont {M.}~\bibnamefont
  {Liu}}, \bibinfo {author} {\bibfnamefont {C.}~\bibnamefont {Wang}}, \bibinfo
  {author} {\bibfnamefont {J.-X.}\ \bibnamefont {Yin}}, \bibinfo {author}
  {\bibfnamefont {J.}~\bibnamefont {Duan}}, \bibinfo {author} {\bibfnamefont
  {X.}~\bibnamefont {Li}}, \bibinfo {author} {\bibfnamefont {Z.}~\bibnamefont
  {Wang}}, \bibinfo {author} {\bibfnamefont {H.-H.}\ \bibnamefont {Wen}},\ and\
  \bibinfo {author} {\bibfnamefont {Y.}~\bibnamefont {Yao}},\ }\href
  {https://doi.org/10.1103/PhysRevB.105.L180507} {\bibfield  {journal}
  {\bibinfo  {journal} {Phys. Rev. B}\ }\textbf {\bibinfo {volume} {105}},\
  \bibinfo {pages} {L180507} (\bibinfo {year}
  {2022}{\natexlab{a}})}\BibitemShut {NoStop}%
\bibitem [{\citenamefont {Jiang}\ \emph {et~al.}(2021)\citenamefont {Jiang},
  \citenamefont {Yin}, \citenamefont {Denner}, \citenamefont {Shumiya},
  \citenamefont {Ortiz}, \citenamefont {Xu}, \citenamefont {Guguchia},
  \citenamefont {He}, \citenamefont {Hossain}, \citenamefont {Liu},
  \citenamefont {Ruff}, \citenamefont {Kautzsch}, \citenamefont {Zhang},
  \citenamefont {Chang}, \citenamefont {Belopolski}, \citenamefont {Zhang},
  \citenamefont {Cochran}, \citenamefont {Multer}, \citenamefont {Litskevich},
  \citenamefont {Cheng}, \citenamefont {Yang}, \citenamefont {Wang},
  \citenamefont {Thomale}, \citenamefont {Neupert}, \citenamefont {Wilson},\
  and\ \citenamefont {Hasan}}]{JiangNM2021}%
  \BibitemOpen
  \bibfield  {author} {\bibinfo {author} {\bibfnamefont {Y.-X.}\ \bibnamefont
  {Jiang}}, \bibinfo {author} {\bibfnamefont {J.-X.}\ \bibnamefont {Yin}},
  \bibinfo {author} {\bibfnamefont {M.~M.}\ \bibnamefont {Denner}}, \bibinfo
  {author} {\bibfnamefont {N.}~\bibnamefont {Shumiya}}, \bibinfo {author}
  {\bibfnamefont {B.~R.}\ \bibnamefont {Ortiz}}, \bibinfo {author}
  {\bibfnamefont {G.}~\bibnamefont {Xu}}, \bibinfo {author} {\bibfnamefont
  {Z.}~\bibnamefont {Guguchia}}, \bibinfo {author} {\bibfnamefont
  {J.}~\bibnamefont {He}}, \bibinfo {author} {\bibfnamefont {M.~S.}\
  \bibnamefont {Hossain}}, \bibinfo {author} {\bibfnamefont {X.}~\bibnamefont
  {Liu}}, \bibinfo {author} {\bibfnamefont {J.}~\bibnamefont {Ruff}}, \bibinfo
  {author} {\bibfnamefont {L.}~\bibnamefont {Kautzsch}}, \bibinfo {author}
  {\bibfnamefont {S.~S.}\ \bibnamefont {Zhang}}, \bibinfo {author}
  {\bibfnamefont {G.}~\bibnamefont {Chang}}, \bibinfo {author} {\bibfnamefont
  {I.}~\bibnamefont {Belopolski}}, \bibinfo {author} {\bibfnamefont
  {Q.}~\bibnamefont {Zhang}}, \bibinfo {author} {\bibfnamefont {T.~A.}\
  \bibnamefont {Cochran}}, \bibinfo {author} {\bibfnamefont {D.}~\bibnamefont
  {Multer}}, \bibinfo {author} {\bibfnamefont {M.}~\bibnamefont {Litskevich}},
  \bibinfo {author} {\bibfnamefont {Z.-J.}\ \bibnamefont {Cheng}}, \bibinfo
  {author} {\bibfnamefont {X.~P.}\ \bibnamefont {Yang}}, \bibinfo {author}
  {\bibfnamefont {Z.}~\bibnamefont {Wang}}, \bibinfo {author} {\bibfnamefont
  {R.}~\bibnamefont {Thomale}}, \bibinfo {author} {\bibfnamefont
  {T.}~\bibnamefont {Neupert}}, \bibinfo {author} {\bibfnamefont {S.~D.}\
  \bibnamefont {Wilson}},\ and\ \bibinfo {author} {\bibfnamefont {M.~Z.}\
  \bibnamefont {Hasan}},\ }\href@noop {} {\bibfield  {journal} {\bibinfo
  {journal} {Nat. Mater.}\ }\textbf {\bibinfo {volume} {20}},\ \bibinfo {pages}
  {1353} (\bibinfo {year} {2021})}\BibitemShut {NoStop}%
\bibitem [{\citenamefont {Mielke~III}\ \emph {et~al.}(2022)\citenamefont
  {Mielke~III}, \citenamefont {Das}, \citenamefont {Yin}, \citenamefont {Liu},
  \citenamefont {Gupta}, \citenamefont {Jiang}, \citenamefont {Medarde},
  \citenamefont {Wu}, \citenamefont {Lei}, \citenamefont {Chang}, \citenamefont
  {Dai}, \citenamefont {Si}, \citenamefont {Miao}, \citenamefont {Thomale},
  \citenamefont {Neupert}, \citenamefont {Shi}, \citenamefont {Khasanov},
  \citenamefont {Hasan}, \citenamefont {Luetkens},\ and\ \citenamefont
  {Guguchia}}]{MielkeNature2022}%
  \BibitemOpen
  \bibfield  {author} {\bibinfo {author} {\bibfnamefont {C.}~\bibnamefont
  {Mielke~III}}, \bibinfo {author} {\bibfnamefont {D.}~\bibnamefont {Das}},
  \bibinfo {author} {\bibfnamefont {J.-X.}\ \bibnamefont {Yin}}, \bibinfo
  {author} {\bibfnamefont {H.}~\bibnamefont {Liu}}, \bibinfo {author}
  {\bibfnamefont {R.}~\bibnamefont {Gupta}}, \bibinfo {author} {\bibfnamefont
  {Y.-X.}\ \bibnamefont {Jiang}}, \bibinfo {author} {\bibfnamefont
  {M.}~\bibnamefont {Medarde}}, \bibinfo {author} {\bibfnamefont
  {X.}~\bibnamefont {Wu}}, \bibinfo {author} {\bibfnamefont {H.~C.}\
  \bibnamefont {Lei}}, \bibinfo {author} {\bibfnamefont {J.}~\bibnamefont
  {Chang}}, \bibinfo {author} {\bibfnamefont {P.}~\bibnamefont {Dai}}, \bibinfo
  {author} {\bibfnamefont {Q.}~\bibnamefont {Si}}, \bibinfo {author}
  {\bibfnamefont {H.}~\bibnamefont {Miao}}, \bibinfo {author} {\bibfnamefont
  {R.}~\bibnamefont {Thomale}}, \bibinfo {author} {\bibfnamefont
  {T.}~\bibnamefont {Neupert}}, \bibinfo {author} {\bibfnamefont
  {Y.}~\bibnamefont {Shi}}, \bibinfo {author} {\bibfnamefont {R.}~\bibnamefont
  {Khasanov}}, \bibinfo {author} {\bibfnamefont {M.~Z.}\ \bibnamefont {Hasan}},
  \bibinfo {author} {\bibfnamefont {H.}~\bibnamefont {Luetkens}},\ and\
  \bibinfo {author} {\bibfnamefont {Z.}~\bibnamefont {Guguchia}},\ }\href@noop
  {} {\bibfield  {journal} {\bibinfo  {journal} {Nature}\ }\textbf {\bibinfo
  {volume} {602}},\ \bibinfo {pages} {245} (\bibinfo {year}
  {2022})}\BibitemShut {NoStop}%
\bibitem [{\citenamefont {Shumiya}\ \emph {et~al.}(2021)\citenamefont
  {Shumiya}, \citenamefont {Hossain}, \citenamefont {Yin}, \citenamefont
  {Jiang}, \citenamefont {Ortiz}, \citenamefont {Liu}, \citenamefont {Shi},
  \citenamefont {Yin}, \citenamefont {Lei}, \citenamefont {Zhang},
  \citenamefont {Chang}, \citenamefont {Zhang}, \citenamefont {Cochran},
  \citenamefont {Multer}, \citenamefont {Litskevich}, \citenamefont {Cheng},
  \citenamefont {Yang}, \citenamefont {Guguchia}, \citenamefont {Wilson},\ and\
  \citenamefont {Hasan}}]{ShumiyaPRB2021}%
  \BibitemOpen
  \bibfield  {author} {\bibinfo {author} {\bibfnamefont {N.}~\bibnamefont
  {Shumiya}}, \bibinfo {author} {\bibfnamefont {M.~S.}\ \bibnamefont
  {Hossain}}, \bibinfo {author} {\bibfnamefont {J.-X.}\ \bibnamefont {Yin}},
  \bibinfo {author} {\bibfnamefont {Y.-X.}\ \bibnamefont {Jiang}}, \bibinfo
  {author} {\bibfnamefont {B.~R.}\ \bibnamefont {Ortiz}}, \bibinfo {author}
  {\bibfnamefont {H.}~\bibnamefont {Liu}}, \bibinfo {author} {\bibfnamefont
  {Y.}~\bibnamefont {Shi}}, \bibinfo {author} {\bibfnamefont {Q.}~\bibnamefont
  {Yin}}, \bibinfo {author} {\bibfnamefont {H.}~\bibnamefont {Lei}}, \bibinfo
  {author} {\bibfnamefont {S.~S.}\ \bibnamefont {Zhang}}, \bibinfo {author}
  {\bibfnamefont {G.}~\bibnamefont {Chang}}, \bibinfo {author} {\bibfnamefont
  {Q.}~\bibnamefont {Zhang}}, \bibinfo {author} {\bibfnamefont {T.~A.}\
  \bibnamefont {Cochran}}, \bibinfo {author} {\bibfnamefont {D.}~\bibnamefont
  {Multer}}, \bibinfo {author} {\bibfnamefont {M.}~\bibnamefont {Litskevich}},
  \bibinfo {author} {\bibfnamefont {Z.-J.}\ \bibnamefont {Cheng}}, \bibinfo
  {author} {\bibfnamefont {X.~P.}\ \bibnamefont {Yang}}, \bibinfo {author}
  {\bibfnamefont {Z.}~\bibnamefont {Guguchia}}, \bibinfo {author}
  {\bibfnamefont {S.~D.}\ \bibnamefont {Wilson}},\ and\ \bibinfo {author}
  {\bibfnamefont {M.~Z.}\ \bibnamefont {Hasan}},\ }\href@noop {} {\bibfield
  {journal} {\bibinfo  {journal} {Phys. Rev. B}\ }\textbf {\bibinfo {volume}
  {104}},\ \bibinfo {pages} {035131} (\bibinfo {year} {2021})}\BibitemShut
  {NoStop}%
\bibitem [{\citenamefont {Wang}\ \emph {et~al.}(2021)\citenamefont {Wang},
  \citenamefont {Jiang}, \citenamefont {Yin}, \citenamefont {Li}, \citenamefont
  {Wang}, \citenamefont {Huang}, \citenamefont {Shao}, \citenamefont {Liu},
  \citenamefont {Zhu}, \citenamefont {Shumiya}, \citenamefont {Hossain},
  \citenamefont {Liu}, \citenamefont {Shi}, \citenamefont {Duan}, \citenamefont
  {Li}, \citenamefont {Chang}, \citenamefont {Dai}, \citenamefont {Ye},
  \citenamefont {Xu}, \citenamefont {Wang}, \citenamefont {Zheng},
  \citenamefont {Jia}, \citenamefont {Hasan},\ and\ \citenamefont
  {Yao}}]{WangPRB2021}%
  \BibitemOpen
  \bibfield  {author} {\bibinfo {author} {\bibfnamefont {Z.}~\bibnamefont
  {Wang}}, \bibinfo {author} {\bibfnamefont {Y.-X.}\ \bibnamefont {Jiang}},
  \bibinfo {author} {\bibfnamefont {J.-X.}\ \bibnamefont {Yin}}, \bibinfo
  {author} {\bibfnamefont {Y.}~\bibnamefont {Li}}, \bibinfo {author}
  {\bibfnamefont {G.-Y.}\ \bibnamefont {Wang}}, \bibinfo {author}
  {\bibfnamefont {H.-L.}\ \bibnamefont {Huang}}, \bibinfo {author}
  {\bibfnamefont {S.}~\bibnamefont {Shao}}, \bibinfo {author} {\bibfnamefont
  {J.}~\bibnamefont {Liu}}, \bibinfo {author} {\bibfnamefont {P.}~\bibnamefont
  {Zhu}}, \bibinfo {author} {\bibfnamefont {N.}~\bibnamefont {Shumiya}},
  \bibinfo {author} {\bibfnamefont {M.~S.}\ \bibnamefont {Hossain}}, \bibinfo
  {author} {\bibfnamefont {H.}~\bibnamefont {Liu}}, \bibinfo {author}
  {\bibfnamefont {Y.}~\bibnamefont {Shi}}, \bibinfo {author} {\bibfnamefont
  {J.}~\bibnamefont {Duan}}, \bibinfo {author} {\bibfnamefont {X.}~\bibnamefont
  {Li}}, \bibinfo {author} {\bibfnamefont {G.}~\bibnamefont {Chang}}, \bibinfo
  {author} {\bibfnamefont {P.}~\bibnamefont {Dai}}, \bibinfo {author}
  {\bibfnamefont {Z.}~\bibnamefont {Ye}}, \bibinfo {author} {\bibfnamefont
  {G.}~\bibnamefont {Xu}}, \bibinfo {author} {\bibfnamefont {Y.}~\bibnamefont
  {Wang}}, \bibinfo {author} {\bibfnamefont {H.}~\bibnamefont {Zheng}},
  \bibinfo {author} {\bibfnamefont {J.}~\bibnamefont {Jia}}, \bibinfo {author}
  {\bibfnamefont {M.~Z.}\ \bibnamefont {Hasan}},\ and\ \bibinfo {author}
  {\bibfnamefont {Y.}~\bibnamefont {Yao}},\ }\href@noop {} {\bibfield
  {journal} {\bibinfo  {journal} {Phys. Rev. B}\ }\textbf {\bibinfo {volume}
  {104}},\ \bibinfo {pages} {075148} (\bibinfo {year} {2021})}\BibitemShut
  {NoStop}%
\bibitem [{\citenamefont {Nie}\ \emph {et~al.}(2022)\citenamefont {Nie},
  \citenamefont {Sun}, \citenamefont {Ma}, \citenamefont {Song}, \citenamefont
  {Zheng}, \citenamefont {Liang}, \citenamefont {Wu}, \citenamefont {Yu},
  \citenamefont {Li}, \citenamefont {Shan}, \citenamefont {Zhao}, \citenamefont
  {Li}, \citenamefont {Kang}, \citenamefont {Wu}, \citenamefont {Zhou},
  \citenamefont {Liu}, \citenamefont {Xiang}, \citenamefont {Ying},
  \citenamefont {Wang}, \citenamefont {Wu},\ and\ \citenamefont
  {Chen}}]{NieNature2022}%
  \BibitemOpen
  \bibfield  {author} {\bibinfo {author} {\bibfnamefont {L.}~\bibnamefont
  {Nie}}, \bibinfo {author} {\bibfnamefont {K.}~\bibnamefont {Sun}}, \bibinfo
  {author} {\bibfnamefont {W.}~\bibnamefont {Ma}}, \bibinfo {author}
  {\bibfnamefont {D.}~\bibnamefont {Song}}, \bibinfo {author} {\bibfnamefont
  {L.}~\bibnamefont {Zheng}}, \bibinfo {author} {\bibfnamefont
  {Z.}~\bibnamefont {Liang}}, \bibinfo {author} {\bibfnamefont
  {P.}~\bibnamefont {Wu}}, \bibinfo {author} {\bibfnamefont {F.}~\bibnamefont
  {Yu}}, \bibinfo {author} {\bibfnamefont {J.}~\bibnamefont {Li}}, \bibinfo
  {author} {\bibfnamefont {M.}~\bibnamefont {Shan}}, \bibinfo {author}
  {\bibfnamefont {D.}~\bibnamefont {Zhao}}, \bibinfo {author} {\bibfnamefont
  {S.}~\bibnamefont {Li}}, \bibinfo {author} {\bibfnamefont {B.}~\bibnamefont
  {Kang}}, \bibinfo {author} {\bibfnamefont {Z.}~\bibnamefont {Wu}}, \bibinfo
  {author} {\bibfnamefont {Y.}~\bibnamefont {Zhou}}, \bibinfo {author}
  {\bibfnamefont {K.}~\bibnamefont {Liu}}, \bibinfo {author} {\bibfnamefont
  {Z.}~\bibnamefont {Xiang}}, \bibinfo {author} {\bibfnamefont
  {J.}~\bibnamefont {Ying}}, \bibinfo {author} {\bibfnamefont {Z.}~\bibnamefont
  {Wang}}, \bibinfo {author} {\bibfnamefont {T.}~\bibnamefont {Wu}},\ and\
  \bibinfo {author} {\bibfnamefont {X.}~\bibnamefont {Chen}},\ }\href@noop {}
  {\bibfield  {journal} {\bibinfo  {journal} {Nature}\ }\textbf {\bibinfo
  {volume} {604}},\ \bibinfo {pages} {59} (\bibinfo {year} {2022})}\BibitemShut
  {NoStop}%
\bibitem [{\citenamefont {Xu}\ \emph {et~al.}(2022)\citenamefont {Xu},
  \citenamefont {Ni}, \citenamefont {Liu}, \citenamefont {Ortiz}, \citenamefont
  {Deng}, \citenamefont {Wilson}, \citenamefont {Yan}, \citenamefont
  {Balents},\ and\ \citenamefont {Wu}}]{XuNP2022}%
  \BibitemOpen
  \bibfield  {author} {\bibinfo {author} {\bibfnamefont {Y.}~\bibnamefont
  {Xu}}, \bibinfo {author} {\bibfnamefont {Z.}~\bibnamefont {Ni}}, \bibinfo
  {author} {\bibfnamefont {Y.}~\bibnamefont {Liu}}, \bibinfo {author}
  {\bibfnamefont {B.~R.}\ \bibnamefont {Ortiz}}, \bibinfo {author}
  {\bibfnamefont {Q.}~\bibnamefont {Deng}}, \bibinfo {author} {\bibfnamefont
  {S.~D.}\ \bibnamefont {Wilson}}, \bibinfo {author} {\bibfnamefont
  {B.}~\bibnamefont {Yan}}, \bibinfo {author} {\bibfnamefont {L.}~\bibnamefont
  {Balents}},\ and\ \bibinfo {author} {\bibfnamefont {L.}~\bibnamefont {Wu}},\
  }\href@noop {} {\bibfield  {journal} {\bibinfo  {journal} {Nat. Phys.}\
  }\textbf {\bibinfo {volume} {18}},\ \bibinfo {pages} {1470} (\bibinfo {year}
  {2022})}\BibitemShut {NoStop}%
\bibitem [{\citenamefont {Jiang}\ \emph {et~al.}()\citenamefont {Jiang},
  \citenamefont {Ma}, \citenamefont {Xia}, \citenamefont {Xiao}, \citenamefont
  {Liu}, \citenamefont {Liu}, \citenamefont {Yang}, \citenamefont {Ding},
  \citenamefont {Huang}, \citenamefont {Liu}, \citenamefont {Qiao},
  \citenamefont {Liu}, \citenamefont {Peng}, \citenamefont {Cho}, \citenamefont
  {Guo}, \citenamefont {Liu},\ and\ \citenamefont {Shen}}]{JiangarXiv2022}%
  \BibitemOpen
  \bibfield  {author} {\bibinfo {author} {\bibfnamefont {Z.}~\bibnamefont
  {Jiang}}, \bibinfo {author} {\bibfnamefont {H.}~\bibnamefont {Ma}}, \bibinfo
  {author} {\bibfnamefont {W.}~\bibnamefont {Xia}}, \bibinfo {author}
  {\bibfnamefont {Q.}~\bibnamefont {Xiao}}, \bibinfo {author} {\bibfnamefont
  {Z.}~\bibnamefont {Liu}}, \bibinfo {author} {\bibfnamefont {Z.}~\bibnamefont
  {Liu}}, \bibinfo {author} {\bibfnamefont {Y.}~\bibnamefont {Yang}}, \bibinfo
  {author} {\bibfnamefont {J.}~\bibnamefont {Ding}}, \bibinfo {author}
  {\bibfnamefont {Z.}~\bibnamefont {Huang}}, \bibinfo {author} {\bibfnamefont
  {J.}~\bibnamefont {Liu}}, \bibinfo {author} {\bibfnamefont {Y.}~\bibnamefont
  {Qiao}}, \bibinfo {author} {\bibfnamefont {J.}~\bibnamefont {Liu}}, \bibinfo
  {author} {\bibfnamefont {Y.}~\bibnamefont {Peng}}, \bibinfo {author}
  {\bibfnamefont {S.}~\bibnamefont {Cho}}, \bibinfo {author} {\bibfnamefont
  {Y.}~\bibnamefont {Guo}}, \bibinfo {author} {\bibfnamefont {J.}~\bibnamefont
  {Liu}},\ and\ \bibinfo {author} {\bibfnamefont {D.}~\bibnamefont {Shen}},\
  }\bibinfo {note} {arXiv:2208.01499}\BibitemShut {NoStop}%
\bibitem [{\citenamefont {Chen}\ \emph
  {et~al.}(2021{\natexlab{b}})\citenamefont {Chen}, \citenamefont {Yang},
  \citenamefont {Hu}, \citenamefont {Zhao}, \citenamefont {Yuan}, \citenamefont
  {Xing}, \citenamefont {Qian}, \citenamefont {Huang}, \citenamefont {Li},
  \citenamefont {Ye}, \citenamefont {Ma}, \citenamefont {Ni}, \citenamefont
  {Zhang}, \citenamefont {Yin}, \citenamefont {Gong}, \citenamefont {Tu},
  \citenamefont {Lei}, \citenamefont {Tan}, \citenamefont {Zhou}, \citenamefont
  {Shen}, \citenamefont {Dong}, \citenamefont {Yan}, \citenamefont {Wang},\
  and\ \citenamefont {Gao}}]{ChenNature2021}%
  \BibitemOpen
  \bibfield  {author} {\bibinfo {author} {\bibfnamefont {H.}~\bibnamefont
  {Chen}}, \bibinfo {author} {\bibfnamefont {H.}~\bibnamefont {Yang}}, \bibinfo
  {author} {\bibfnamefont {B.}~\bibnamefont {Hu}}, \bibinfo {author}
  {\bibfnamefont {Z.}~\bibnamefont {Zhao}}, \bibinfo {author} {\bibfnamefont
  {J.}~\bibnamefont {Yuan}}, \bibinfo {author} {\bibfnamefont {Y.}~\bibnamefont
  {Xing}}, \bibinfo {author} {\bibfnamefont {G.}~\bibnamefont {Qian}}, \bibinfo
  {author} {\bibfnamefont {Z.}~\bibnamefont {Huang}}, \bibinfo {author}
  {\bibfnamefont {G.}~\bibnamefont {Li}}, \bibinfo {author} {\bibfnamefont
  {Y.}~\bibnamefont {Ye}}, \bibinfo {author} {\bibfnamefont {S.}~\bibnamefont
  {Ma}}, \bibinfo {author} {\bibfnamefont {S.}~\bibnamefont {Ni}}, \bibinfo
  {author} {\bibfnamefont {H.}~\bibnamefont {Zhang}}, \bibinfo {author}
  {\bibfnamefont {Q.}~\bibnamefont {Yin}}, \bibinfo {author} {\bibfnamefont
  {C.}~\bibnamefont {Gong}}, \bibinfo {author} {\bibfnamefont {Z.}~\bibnamefont
  {Tu}}, \bibinfo {author} {\bibfnamefont {H.}~\bibnamefont {Lei}}, \bibinfo
  {author} {\bibfnamefont {H.}~\bibnamefont {Tan}}, \bibinfo {author}
  {\bibfnamefont {S.}~\bibnamefont {Zhou}}, \bibinfo {author} {\bibfnamefont
  {C.}~\bibnamefont {Shen}}, \bibinfo {author} {\bibfnamefont {X.}~\bibnamefont
  {Dong}}, \bibinfo {author} {\bibfnamefont {B.}~\bibnamefont {Yan}}, \bibinfo
  {author} {\bibfnamefont {Z.}~\bibnamefont {Wang}},\ and\ \bibinfo {author}
  {\bibfnamefont {H.-J.}\ \bibnamefont {Gao}},\ }\href@noop {} {\bibfield
  {journal} {\bibinfo  {journal} {Nature}\ }\textbf {\bibinfo {volume} {599}},\
  \bibinfo {pages} {222} (\bibinfo {year} {2021}{\natexlab{b}})}\BibitemShut
  {NoStop}%
\bibitem [{\citenamefont {Zhao}\ \emph {et~al.}(2021)\citenamefont {Zhao},
  \citenamefont {Li}, \citenamefont {Ortiz}, \citenamefont {Teicher},
  \citenamefont {Park}, \citenamefont {Ye}, \citenamefont {Wang}, \citenamefont
  {Balents}, \citenamefont {Wilson},\ and\ \citenamefont
  {Zeljkovic}}]{ZhaoNature2021}%
  \BibitemOpen
  \bibfield  {author} {\bibinfo {author} {\bibfnamefont {H.}~\bibnamefont
  {Zhao}}, \bibinfo {author} {\bibfnamefont {H.}~\bibnamefont {Li}}, \bibinfo
  {author} {\bibfnamefont {B.~R.}\ \bibnamefont {Ortiz}}, \bibinfo {author}
  {\bibfnamefont {S.~M.~L.}\ \bibnamefont {Teicher}}, \bibinfo {author}
  {\bibfnamefont {T.}~\bibnamefont {Park}}, \bibinfo {author} {\bibfnamefont
  {M.}~\bibnamefont {Ye}}, \bibinfo {author} {\bibfnamefont {Z.}~\bibnamefont
  {Wang}}, \bibinfo {author} {\bibfnamefont {L.}~\bibnamefont {Balents}},
  \bibinfo {author} {\bibfnamefont {S.~D.}\ \bibnamefont {Wilson}},\ and\
  \bibinfo {author} {\bibfnamefont {I.}~\bibnamefont {Zeljkovic}},\ }\href@noop
  {} {\bibfield  {journal} {\bibinfo  {journal} {Nature}\ }\textbf {\bibinfo
  {volume} {599}},\ \bibinfo {pages} {216} (\bibinfo {year}
  {2021})}\BibitemShut {NoStop}%
\bibitem [{\citenamefont {Yu}\ \emph {et~al.}(2022)\citenamefont {Yu},
  \citenamefont {Xu}, \citenamefont {Xiao}, \citenamefont {Yuan}, \citenamefont
  {Yin}, \citenamefont {Hu}, \citenamefont {Gong}, \citenamefont {Guo},
  \citenamefont {Tu}, \citenamefont {Tang}, \citenamefont {Lei}, \citenamefont
  {Xue},\ and\ \citenamefont {Li}}]{YuNanoLett2022}%
  \BibitemOpen
  \bibfield  {author} {\bibinfo {author} {\bibfnamefont {J.}~\bibnamefont
  {Yu}}, \bibinfo {author} {\bibfnamefont {Z.}~\bibnamefont {Xu}}, \bibinfo
  {author} {\bibfnamefont {K.}~\bibnamefont {Xiao}}, \bibinfo {author}
  {\bibfnamefont {Y.}~\bibnamefont {Yuan}}, \bibinfo {author} {\bibfnamefont
  {Q.}~\bibnamefont {Yin}}, \bibinfo {author} {\bibfnamefont {Z.}~\bibnamefont
  {Hu}}, \bibinfo {author} {\bibfnamefont {C.}~\bibnamefont {Gong}}, \bibinfo
  {author} {\bibfnamefont {Y.}~\bibnamefont {Guo}}, \bibinfo {author}
  {\bibfnamefont {Z.}~\bibnamefont {Tu}}, \bibinfo {author} {\bibfnamefont
  {P.}~\bibnamefont {Tang}}, \bibinfo {author} {\bibfnamefont {H.}~\bibnamefont
  {Lei}}, \bibinfo {author} {\bibfnamefont {Q.-K.}\ \bibnamefont {Xue}},\ and\
  \bibinfo {author} {\bibfnamefont {W.}~\bibnamefont {Li}},\ }\href@noop {}
  {\bibfield  {journal} {\bibinfo  {journal} {Nano Lett.}\ }\textbf {\bibinfo
  {volume} {22}},\ \bibinfo {pages} {918} (\bibinfo {year} {2022})}\BibitemShut
  {NoStop}%
\bibitem [{\citenamefont {Kato}\ \emph
  {et~al.}(2022{\natexlab{b}})\citenamefont {Kato}, \citenamefont {Li},
  \citenamefont {Nakayama}, \citenamefont {Wang}, \citenamefont {Souma},
  \citenamefont {Kitamura}, \citenamefont {Horiba}, \citenamefont
  {Kumigashira}, \citenamefont {Takahashi},\ and\ \citenamefont
  {Sato}}]{KatoPRB2022}%
  \BibitemOpen
  \bibfield  {author} {\bibinfo {author} {\bibfnamefont {T.}~\bibnamefont
  {Kato}}, \bibinfo {author} {\bibfnamefont {Y.}~\bibnamefont {Li}}, \bibinfo
  {author} {\bibfnamefont {K.}~\bibnamefont {Nakayama}}, \bibinfo {author}
  {\bibfnamefont {Z.}~\bibnamefont {Wang}}, \bibinfo {author} {\bibfnamefont
  {S.}~\bibnamefont {Souma}}, \bibinfo {author} {\bibfnamefont
  {M.}~\bibnamefont {Kitamura}}, \bibinfo {author} {\bibfnamefont
  {K.}~\bibnamefont {Horiba}}, \bibinfo {author} {\bibfnamefont
  {H.}~\bibnamefont {Kumigashira}}, \bibinfo {author} {\bibfnamefont
  {T.}~\bibnamefont {Takahashi}},\ and\ \bibinfo {author} {\bibfnamefont
  {T.}~\bibnamefont {Sato}},\ }\href
  {https://doi.org/10.1103/PhysRevB.106.L121112} {\bibfield  {journal}
  {\bibinfo  {journal} {Phys. Rev. B}\ }\textbf {\bibinfo {volume} {106}},\
  \bibinfo {pages} {L121112} (\bibinfo {year}
  {2022}{\natexlab{b}})}\BibitemShut {NoStop}%
\bibitem [{\citenamefont {Kang}\ \emph
  {et~al.}(2022{\natexlab{a}})\citenamefont {Kang}, \citenamefont {Fang},
  \citenamefont {Yoo}, \citenamefont {Ortiz}, \citenamefont {Oey},
  \citenamefont {Choi}, \citenamefont {Ryu}, \citenamefont {Kim}, \citenamefont
  {Jozwiak}, \citenamefont {Bostwick}, \citenamefont {Rotenberg}, \citenamefont
  {Kaxiras}, \citenamefont {Checkelsky}, \citenamefont {Wilson}, \citenamefont
  {Park},\ and\ \citenamefont {Comin}}]{KangNM2022}%
  \BibitemOpen
  \bibfield  {author} {\bibinfo {author} {\bibfnamefont {M.}~\bibnamefont
  {Kang}}, \bibinfo {author} {\bibfnamefont {S.}~\bibnamefont {Fang}}, \bibinfo
  {author} {\bibfnamefont {J.}~\bibnamefont {Yoo}}, \bibinfo {author}
  {\bibfnamefont {B.~R.}\ \bibnamefont {Ortiz}}, \bibinfo {author}
  {\bibfnamefont {Y.~M.}\ \bibnamefont {Oey}}, \bibinfo {author} {\bibfnamefont
  {J.}~\bibnamefont {Choi}}, \bibinfo {author} {\bibfnamefont {S.~H.}\
  \bibnamefont {Ryu}}, \bibinfo {author} {\bibfnamefont {J.}~\bibnamefont
  {Kim}}, \bibinfo {author} {\bibfnamefont {C.}~\bibnamefont {Jozwiak}},
  \bibinfo {author} {\bibfnamefont {A.}~\bibnamefont {Bostwick}}, \bibinfo
  {author} {\bibfnamefont {E.}~\bibnamefont {Rotenberg}}, \bibinfo {author}
  {\bibfnamefont {E.}~\bibnamefont {Kaxiras}}, \bibinfo {author} {\bibfnamefont
  {J.~G.}\ \bibnamefont {Checkelsky}}, \bibinfo {author} {\bibfnamefont
  {S.~D.}\ \bibnamefont {Wilson}}, \bibinfo {author} {\bibfnamefont {J.-H.}\
  \bibnamefont {Park}},\ and\ \bibinfo {author} {\bibfnamefont
  {R.}~\bibnamefont {Comin}},\ }\href@noop {} {\bibfield  {journal} {\bibinfo
  {journal} {Nat. Mater.}\ }(\bibinfo {year} {2022}{\natexlab{a}})},\ \bibinfo
  {note} {https://doi.org/10.1038/s41563-022-01375-2}\BibitemShut {NoStop}%
\bibitem [{\citenamefont {Li}\ \emph {et~al.}(2022{\natexlab{b}})\citenamefont
  {Li}, \citenamefont {Zhao}, \citenamefont {Ortiz}, \citenamefont {Park},
  \citenamefont {Ye}, \citenamefont {Balents}, \citenamefont {Wang},
  \citenamefont {Wilson},\ and\ \citenamefont {Zeljkovic}}]{LiNP2022}%
  \BibitemOpen
  \bibfield  {author} {\bibinfo {author} {\bibfnamefont {H.}~\bibnamefont
  {Li}}, \bibinfo {author} {\bibfnamefont {H.}~\bibnamefont {Zhao}}, \bibinfo
  {author} {\bibfnamefont {B.~R.}\ \bibnamefont {Ortiz}}, \bibinfo {author}
  {\bibfnamefont {T.}~\bibnamefont {Park}}, \bibinfo {author} {\bibfnamefont
  {M.}~\bibnamefont {Ye}}, \bibinfo {author} {\bibfnamefont {L.}~\bibnamefont
  {Balents}}, \bibinfo {author} {\bibfnamefont {Z.}~\bibnamefont {Wang}},
  \bibinfo {author} {\bibfnamefont {S.~D.}\ \bibnamefont {Wilson}},\ and\
  \bibinfo {author} {\bibfnamefont {I.}~\bibnamefont {Zeljkovic}},\ }\href@noop
  {} {\bibfield  {journal} {\bibinfo  {journal} {Nat. Phys.}\ }\textbf
  {\bibinfo {volume} {18}},\ \bibinfo {pages} {265} (\bibinfo {year}
  {2022}{\natexlab{b}})}\BibitemShut {NoStop}%
\bibitem [{\citenamefont {Kautzsch}\ \emph {et~al.}()\citenamefont {Kautzsch},
  \citenamefont {Ortiz}, \citenamefont {Mallayya}, \citenamefont {Plumb},
  \citenamefont {Pokharel}, \citenamefont {Ruff}, \citenamefont {Islam},
  \citenamefont {Kim}, \citenamefont {Seshadri},\ and\ \citenamefont
  {Wilson}}]{KautzscharXiv2022}%
  \BibitemOpen
  \bibfield  {author} {\bibinfo {author} {\bibfnamefont {L.}~\bibnamefont
  {Kautzsch}}, \bibinfo {author} {\bibfnamefont {B.~R.}\ \bibnamefont {Ortiz}},
  \bibinfo {author} {\bibfnamefont {K.}~\bibnamefont {Mallayya}}, \bibinfo
  {author} {\bibfnamefont {J.}~\bibnamefont {Plumb}}, \bibinfo {author}
  {\bibfnamefont {G.}~\bibnamefont {Pokharel}}, \bibinfo {author}
  {\bibfnamefont {J.~P.~C.}\ \bibnamefont {Ruff}}, \bibinfo {author}
  {\bibfnamefont {Z.}~\bibnamefont {Islam}}, \bibinfo {author} {\bibfnamefont
  {E.-A.}\ \bibnamefont {Kim}}, \bibinfo {author} {\bibfnamefont
  {R.}~\bibnamefont {Seshadri}},\ and\ \bibinfo {author} {\bibfnamefont
  {S.~D.}\ \bibnamefont {Wilson}},\ }\bibinfo {note}
  {arXiv:2211.16602}\BibitemShut {NoStop}%
\bibitem [{\citenamefont {Kitamura}\ \emph {et~al.}(2022)\citenamefont
  {Kitamura}, \citenamefont {Souma}, \citenamefont {Honma}, \citenamefont
  {Wakabayashi}, \citenamefont {Tanaka}, \citenamefont {Toyoshima},
  \citenamefont {Amemiya}, \citenamefont {Kawakami}, \citenamefont {Sugawara},
  \citenamefont {Nakayama}, \citenamefont {Yoshimatsu}, \citenamefont
  {Kumigashira}, \citenamefont {Sato},\ and\ \citenamefont
  {Horiba}}]{KitamuraRSI2022}%
  \BibitemOpen
  \bibfield  {author} {\bibinfo {author} {\bibfnamefont {M.}~\bibnamefont
  {Kitamura}}, \bibinfo {author} {\bibfnamefont {S.}~\bibnamefont {Souma}},
  \bibinfo {author} {\bibfnamefont {A.}~\bibnamefont {Honma}}, \bibinfo
  {author} {\bibfnamefont {D.}~\bibnamefont {Wakabayashi}}, \bibinfo {author}
  {\bibfnamefont {H.}~\bibnamefont {Tanaka}}, \bibinfo {author} {\bibfnamefont
  {A.}~\bibnamefont {Toyoshima}}, \bibinfo {author} {\bibfnamefont
  {K.}~\bibnamefont {Amemiya}}, \bibinfo {author} {\bibfnamefont
  {T.}~\bibnamefont {Kawakami}}, \bibinfo {author} {\bibfnamefont
  {K.}~\bibnamefont {Sugawara}}, \bibinfo {author} {\bibfnamefont
  {K.}~\bibnamefont {Nakayama}}, \bibinfo {author} {\bibfnamefont
  {K.}~\bibnamefont {Yoshimatsu}}, \bibinfo {author} {\bibfnamefont
  {H.}~\bibnamefont {Kumigashira}}, \bibinfo {author} {\bibfnamefont
  {T.}~\bibnamefont {Sato}},\ and\ \bibinfo {author} {\bibfnamefont
  {K.}~\bibnamefont {Horiba}},\ }\href@noop {} {\bibfield  {journal} {\bibinfo
  {journal} {Rev. Sci. Instrum.}\ }\textbf {\bibinfo {volume} {93}},\ \bibinfo
  {pages} {033906} (\bibinfo {year} {2022})}\BibitemShut {NoStop}%
\bibitem [{\citenamefont {Kresse}\ and\ \citenamefont
  {Furthm\"uller}(1996{\natexlab{a}})}]{KressePRB1996}%
  \BibitemOpen
  \bibfield  {author} {\bibinfo {author} {\bibfnamefont {G.}~\bibnamefont
  {Kresse}}\ and\ \bibinfo {author} {\bibfnamefont {J.}~\bibnamefont
  {Furthm\"uller}},\ }\href {https://doi.org/10.1103/PhysRevB.54.11169}
  {\bibfield  {journal} {\bibinfo  {journal} {Phys. Rev. B}\ }\textbf {\bibinfo
  {volume} {54}},\ \bibinfo {pages} {11169} (\bibinfo {year}
  {1996}{\natexlab{a}})}\BibitemShut {NoStop}%
\bibitem [{\citenamefont {Kresse}\ and\ \citenamefont
  {Furthm\"uller}(1996{\natexlab{b}})}]{KresseCMS1996}%
  \BibitemOpen
  \bibfield  {author} {\bibinfo {author} {\bibfnamefont {G.}~\bibnamefont
  {Kresse}}\ and\ \bibinfo {author} {\bibfnamefont {J.}~\bibnamefont
  {Furthm\"uller}},\ }\href
  {https://doi.org/https://doi.org/10.1016/0927-0256(96)00008-0} {\bibfield
  {journal} {\bibinfo  {journal} {Comput. Mater. Sci.}\ }\textbf {\bibinfo
  {volume} {6}},\ \bibinfo {pages} {15} (\bibinfo {year}
  {1996}{\natexlab{b}})}\BibitemShut {NoStop}%
\bibitem [{\citenamefont {Bl\"ochl}(1994)}]{BlochlPRB1994}%
  \BibitemOpen
  \bibfield  {author} {\bibinfo {author} {\bibfnamefont {P.~E.}\ \bibnamefont
  {Bl\"ochl}},\ }\href {https://doi.org/10.1103/PhysRevB.50.17953} {\bibfield
  {journal} {\bibinfo  {journal} {Phys. Rev. B}\ }\textbf {\bibinfo {volume}
  {50}},\ \bibinfo {pages} {17953} (\bibinfo {year} {1994})}\BibitemShut
  {NoStop}%
\bibitem [{\citenamefont {Perdew}\ \emph {et~al.}(1996)\citenamefont {Perdew},
  \citenamefont {Burke},\ and\ \citenamefont {Ernzerhof}}]{PerdewPRL1996}%
  \BibitemOpen
  \bibfield  {author} {\bibinfo {author} {\bibfnamefont {J.~P.}\ \bibnamefont
  {Perdew}}, \bibinfo {author} {\bibfnamefont {K.}~\bibnamefont {Burke}},\ and\
  \bibinfo {author} {\bibfnamefont {M.}~\bibnamefont {Ernzerhof}},\ }\href
  {https://doi.org/10.1103/PhysRevLett.77.3865} {\bibfield  {journal} {\bibinfo
   {journal} {Phys. Rev. Lett.}\ }\textbf {\bibinfo {volume} {77}},\ \bibinfo
  {pages} {3865} (\bibinfo {year} {1996})}\BibitemShut {NoStop}%
\bibitem [{\citenamefont {Grimme}\ \emph {et~al.}(2010)\citenamefont {Grimme},
  \citenamefont {Antony}, \citenamefont {Ehrlich},\ and\ \citenamefont
  {Krieg}}]{GrimmeJCP2010}%
  \BibitemOpen
  \bibfield  {author} {\bibinfo {author} {\bibfnamefont {S.}~\bibnamefont
  {Grimme}}, \bibinfo {author} {\bibfnamefont {J.}~\bibnamefont {Antony}},
  \bibinfo {author} {\bibfnamefont {S.}~\bibnamefont {Ehrlich}},\ and\ \bibinfo
  {author} {\bibfnamefont {H.}~\bibnamefont {Krieg}},\ }\href@noop {}
  {\bibfield  {journal} {\bibinfo  {journal} {J. Chem. Phys.}\ }\textbf
  {\bibinfo {volume} {132}},\ \bibinfo {pages} {154104} (\bibinfo {year}
  {2010})}\BibitemShut {NoStop}%
\bibitem [{\citenamefont {Wu}\ \emph {et~al.}(2018)\citenamefont {Wu},
  \citenamefont {Zhang}, \citenamefont {Song}, \citenamefont {Troyer},\ and\
  \citenamefont {Soluyanov}}]{WuCPC2018}%
  \BibitemOpen
  \bibfield  {author} {\bibinfo {author} {\bibfnamefont {Q.}~\bibnamefont
  {Wu}}, \bibinfo {author} {\bibfnamefont {S.}~\bibnamefont {Zhang}}, \bibinfo
  {author} {\bibfnamefont {H.-F.}\ \bibnamefont {Song}}, \bibinfo {author}
  {\bibfnamefont {M.}~\bibnamefont {Troyer}},\ and\ \bibinfo {author}
  {\bibfnamefont {A.~A.}\ \bibnamefont {Soluyanov}},\ }\href
  {https://doi.org/https://doi.org/10.1016/j.cpc.2017.09.033} {\bibfield
  {journal} {\bibinfo  {journal} {Comput. Phys. Commun.}\ }\textbf {\bibinfo
  {volume} {224}},\ \bibinfo {pages} {405} (\bibinfo {year}
  {2018})}\BibitemShut {NoStop}%
\bibitem [{\citenamefont {Nakayama}\ \emph {et~al.}(2021)\citenamefont
  {Nakayama}, \citenamefont {Li}, \citenamefont {Kato}, \citenamefont {Liu},
  \citenamefont {Wang}, \citenamefont {Takahashi}, \citenamefont {Yao},\ and\
  \citenamefont {Sato}}]{NakayamaPRB2021}%
  \BibitemOpen
  \bibfield  {author} {\bibinfo {author} {\bibfnamefont {K.}~\bibnamefont
  {Nakayama}}, \bibinfo {author} {\bibfnamefont {Y.}~\bibnamefont {Li}},
  \bibinfo {author} {\bibfnamefont {T.}~\bibnamefont {Kato}}, \bibinfo {author}
  {\bibfnamefont {M.}~\bibnamefont {Liu}}, \bibinfo {author} {\bibfnamefont
  {Z.}~\bibnamefont {Wang}}, \bibinfo {author} {\bibfnamefont {T.}~\bibnamefont
  {Takahashi}}, \bibinfo {author} {\bibfnamefont {Y.}~\bibnamefont {Yao}},\
  and\ \bibinfo {author} {\bibfnamefont {T.}~\bibnamefont {Sato}},\ }\href@noop
  {} {\bibfield  {journal} {\bibinfo  {journal} {Phys. Rev. B}\ }\textbf
  {\bibinfo {volume} {104}},\ \bibinfo {pages} {L161112} (\bibinfo {year}
  {2021})}\BibitemShut {NoStop}%
\bibitem [{\citenamefont {Wang}\ \emph {et~al.}(2022)\citenamefont {Wang},
  \citenamefont {Liu}, \citenamefont {Jeon},\ and\ \citenamefont
  {Cho}}]{WangPRB2022}%
  \BibitemOpen
  \bibfield  {author} {\bibinfo {author} {\bibfnamefont {C.}~\bibnamefont
  {Wang}}, \bibinfo {author} {\bibfnamefont {S.}~\bibnamefont {Liu}}, \bibinfo
  {author} {\bibfnamefont {H.}~\bibnamefont {Jeon}},\ and\ \bibinfo {author}
  {\bibfnamefont {J.-H.}\ \bibnamefont {Cho}},\ }\href@noop {} {\bibfield
  {journal} {\bibinfo  {journal} {Phys. Rev. B}\ }\textbf {\bibinfo {volume}
  {105}},\ \bibinfo {pages} {045135} (\bibinfo {year} {2022})}\BibitemShut
  {NoStop}%
\bibitem [{\citenamefont {Tan}\ \emph {et~al.}(2021)\citenamefont {Tan},
  \citenamefont {Liu}, \citenamefont {Wang},\ and\ \citenamefont
  {Yan}}]{TanPRL2021}%
  \BibitemOpen
  \bibfield  {author} {\bibinfo {author} {\bibfnamefont {H.}~\bibnamefont
  {Tan}}, \bibinfo {author} {\bibfnamefont {Y.}~\bibnamefont {Liu}}, \bibinfo
  {author} {\bibfnamefont {Z.}~\bibnamefont {Wang}},\ and\ \bibinfo {author}
  {\bibfnamefont {B.}~\bibnamefont {Yan}},\ }\href@noop {} {\bibfield
  {journal} {\bibinfo  {journal} {Phys. Rev. Lett.}\ }\textbf {\bibinfo
  {volume} {127}},\ \bibinfo {pages} {046401} (\bibinfo {year}
  {2021})}\BibitemShut {NoStop}%
\bibitem [{\citenamefont {Fu}\ \emph {et~al.}(2021)\citenamefont {Fu},
  \citenamefont {Zhao}, \citenamefont {Chen}, \citenamefont {Yin},
  \citenamefont {Tu}, \citenamefont {Gong}, \citenamefont {Xi}, \citenamefont
  {Zhu}, \citenamefont {Sun}, \citenamefont {Liu},\ and\ \citenamefont
  {Lei}}]{FuPRL2021}%
  \BibitemOpen
  \bibfield  {author} {\bibinfo {author} {\bibfnamefont {Y.}~\bibnamefont
  {Fu}}, \bibinfo {author} {\bibfnamefont {N.}~\bibnamefont {Zhao}}, \bibinfo
  {author} {\bibfnamefont {Z.}~\bibnamefont {Chen}}, \bibinfo {author}
  {\bibfnamefont {Q.}~\bibnamefont {Yin}}, \bibinfo {author} {\bibfnamefont
  {Z.}~\bibnamefont {Tu}}, \bibinfo {author} {\bibfnamefont {C.}~\bibnamefont
  {Gong}}, \bibinfo {author} {\bibfnamefont {C.}~\bibnamefont {Xi}}, \bibinfo
  {author} {\bibfnamefont {X.}~\bibnamefont {Zhu}}, \bibinfo {author}
  {\bibfnamefont {Y.}~\bibnamefont {Sun}}, \bibinfo {author} {\bibfnamefont
  {K.}~\bibnamefont {Liu}},\ and\ \bibinfo {author} {\bibfnamefont
  {H.}~\bibnamefont {Lei}},\ }\href@noop {} {\bibfield  {journal} {\bibinfo
  {journal} {Phys. Rev. Lett.}\ }\textbf {\bibinfo {volume} {127}},\ \bibinfo
  {pages} {207002} (\bibinfo {year} {2021})}\BibitemShut {NoStop}%
\bibitem [{\citenamefont {Kang}\ \emph
  {et~al.}(2022{\natexlab{b}})\citenamefont {Kang}, \citenamefont {Fang},
  \citenamefont {Kim}, \citenamefont {Ortiz}, \citenamefont {Ryu},
  \citenamefont {Kim}, \citenamefont {Yoo}, \citenamefont {Sangiovanni},
  \citenamefont {Sante}, \citenamefont {Park}, \citenamefont {Jozwiak},
  \citenamefont {Bostwick}, \citenamefont {Rotenberg}, \citenamefont {Kaxiras},
  \citenamefont {Wilson}, \citenamefont {Park},\ and\ \citenamefont
  {Comin}}]{KangNP2022}%
  \BibitemOpen
  \bibfield  {author} {\bibinfo {author} {\bibfnamefont {M.}~\bibnamefont
  {Kang}}, \bibinfo {author} {\bibfnamefont {S.}~\bibnamefont {Fang}}, \bibinfo
  {author} {\bibfnamefont {J.-K.}\ \bibnamefont {Kim}}, \bibinfo {author}
  {\bibfnamefont {B.~R.}\ \bibnamefont {Ortiz}}, \bibinfo {author}
  {\bibfnamefont {S.~H.}\ \bibnamefont {Ryu}}, \bibinfo {author} {\bibfnamefont
  {J.}~\bibnamefont {Kim}}, \bibinfo {author} {\bibfnamefont {J.}~\bibnamefont
  {Yoo}}, \bibinfo {author} {\bibfnamefont {G.}~\bibnamefont {Sangiovanni}},
  \bibinfo {author} {\bibfnamefont {D.~D.}\ \bibnamefont {Sante}}, \bibinfo
  {author} {\bibfnamefont {B.-G.}\ \bibnamefont {Park}}, \bibinfo {author}
  {\bibfnamefont {C.}~\bibnamefont {Jozwiak}}, \bibinfo {author} {\bibfnamefont
  {A.}~\bibnamefont {Bostwick}}, \bibinfo {author} {\bibfnamefont
  {E.}~\bibnamefont {Rotenberg}}, \bibinfo {author} {\bibfnamefont
  {E.}~\bibnamefont {Kaxiras}}, \bibinfo {author} {\bibfnamefont {S.~D.}\
  \bibnamefont {Wilson}}, \bibinfo {author} {\bibfnamefont {J.-H.}\
  \bibnamefont {Park}},\ and\ \bibinfo {author} {\bibfnamefont
  {R.}~\bibnamefont {Comin}},\ }\href@noop {} {\bibfield  {journal} {\bibinfo
  {journal} {Nat. Phys.}\ }\textbf {\bibinfo {volume} {18}},\ \bibinfo {pages}
  {301} (\bibinfo {year} {2022}{\natexlab{b}})}\BibitemShut {NoStop}%
\bibitem [{\citenamefont {Hu}\ \emph {et~al.}(2022{\natexlab{a}})\citenamefont
  {Hu}, \citenamefont {Wu}, \citenamefont {Ortiz}, \citenamefont {Ju},
  \citenamefont {Han}, \citenamefont {Ma}, \citenamefont {Plumb}, \citenamefont
  {Radovic}, \citenamefont {Thomale}, \citenamefont {Wilson}, \citenamefont
  {Schnyder},\ and\ \citenamefont {Shi}}]{HuNCOM2022}%
  \BibitemOpen
  \bibfield  {author} {\bibinfo {author} {\bibfnamefont {Y.}~\bibnamefont
  {Hu}}, \bibinfo {author} {\bibfnamefont {X.}~\bibnamefont {Wu}}, \bibinfo
  {author} {\bibfnamefont {B.~R.}\ \bibnamefont {Ortiz}}, \bibinfo {author}
  {\bibfnamefont {S.}~\bibnamefont {Ju}}, \bibinfo {author} {\bibfnamefont
  {X.}~\bibnamefont {Han}}, \bibinfo {author} {\bibfnamefont {J.}~\bibnamefont
  {Ma}}, \bibinfo {author} {\bibfnamefont {N.~C.}\ \bibnamefont {Plumb}},
  \bibinfo {author} {\bibfnamefont {M.}~\bibnamefont {Radovic}}, \bibinfo
  {author} {\bibfnamefont {R.}~\bibnamefont {Thomale}}, \bibinfo {author}
  {\bibfnamefont {S.~D.}\ \bibnamefont {Wilson}}, \bibinfo {author}
  {\bibfnamefont {A.~P.}\ \bibnamefont {Schnyder}},\ and\ \bibinfo {author}
  {\bibfnamefont {M.}~\bibnamefont {Shi}},\ }\href@noop {} {\bibfield
  {journal} {\bibinfo  {journal} {Nat. Commun.}\ }\textbf {\bibinfo {volume}
  {13}},\ \bibinfo {pages} {2220} (\bibinfo {year}
  {2022}{\natexlab{a}})}\BibitemShut {NoStop}%
\bibitem [{\citenamefont {Luo}\ \emph {et~al.}(2022)\citenamefont {Luo},
  \citenamefont {Gao}, \citenamefont {Liu}, \citenamefont {Gu}, \citenamefont
  {Wu}, \citenamefont {Yi}, \citenamefont {Jia}, \citenamefont {Wu},
  \citenamefont {Luo}, \citenamefont {Xu}, \citenamefont {Zhao}, \citenamefont
  {Wang}, \citenamefont {Mao}, \citenamefont {Liu}, \citenamefont {Zhu},
  \citenamefont {Shi}, \citenamefont {Jiang}, \citenamefont {Hu}, \citenamefont
  {Xu},\ and\ \citenamefont {Zhou}}]{LuoNCOM2022}%
  \BibitemOpen
  \bibfield  {author} {\bibinfo {author} {\bibfnamefont {H.}~\bibnamefont
  {Luo}}, \bibinfo {author} {\bibfnamefont {Q.}~\bibnamefont {Gao}}, \bibinfo
  {author} {\bibfnamefont {H.}~\bibnamefont {Liu}}, \bibinfo {author}
  {\bibfnamefont {Y.}~\bibnamefont {Gu}}, \bibinfo {author} {\bibfnamefont
  {D.}~\bibnamefont {Wu}}, \bibinfo {author} {\bibfnamefont {C.}~\bibnamefont
  {Yi}}, \bibinfo {author} {\bibfnamefont {J.}~\bibnamefont {Jia}}, \bibinfo
  {author} {\bibfnamefont {S.}~\bibnamefont {Wu}}, \bibinfo {author}
  {\bibfnamefont {X.}~\bibnamefont {Luo}}, \bibinfo {author} {\bibfnamefont
  {Y.}~\bibnamefont {Xu}}, \bibinfo {author} {\bibfnamefont {L.}~\bibnamefont
  {Zhao}}, \bibinfo {author} {\bibfnamefont {Q.}~\bibnamefont {Wang}}, \bibinfo
  {author} {\bibfnamefont {H.}~\bibnamefont {Mao}}, \bibinfo {author}
  {\bibfnamefont {G.}~\bibnamefont {Liu}}, \bibinfo {author} {\bibfnamefont
  {Z.}~\bibnamefont {Zhu}}, \bibinfo {author} {\bibfnamefont {Y.}~\bibnamefont
  {Shi}}, \bibinfo {author} {\bibfnamefont {K.}~\bibnamefont {Jiang}}, \bibinfo
  {author} {\bibfnamefont {J.}~\bibnamefont {Hu}}, \bibinfo {author}
  {\bibfnamefont {Z.}~\bibnamefont {Xu}},\ and\ \bibinfo {author}
  {\bibfnamefont {X.~J.}\ \bibnamefont {Zhou}},\ }\href@noop {} {\bibfield
  {journal} {\bibinfo  {journal} {Nat. Commun.}\ }\textbf {\bibinfo {volume}
  {13}},\ \bibinfo {pages} {273} (\bibinfo {year} {2022})}\BibitemShut
  {NoStop}%
\bibitem [{\citenamefont {Liu}\ \emph {et~al.}(2021)\citenamefont {Liu},
  \citenamefont {Zhao}, \citenamefont {Yin}, \citenamefont {Gong},
  \citenamefont {Tu}, \citenamefont {Li}, \citenamefont {Song}, \citenamefont
  {Liu}, \citenamefont {Shen}, \citenamefont {Huang}, \citenamefont {Liu},
  \citenamefont {Lei},\ and\ \citenamefont {Wang}}]{LiuPRX2021}%
  \BibitemOpen
  \bibfield  {author} {\bibinfo {author} {\bibfnamefont {Z.}~\bibnamefont
  {Liu}}, \bibinfo {author} {\bibfnamefont {N.}~\bibnamefont {Zhao}}, \bibinfo
  {author} {\bibfnamefont {Q.}~\bibnamefont {Yin}}, \bibinfo {author}
  {\bibfnamefont {C.}~\bibnamefont {Gong}}, \bibinfo {author} {\bibfnamefont
  {Z.}~\bibnamefont {Tu}}, \bibinfo {author} {\bibfnamefont {M.}~\bibnamefont
  {Li}}, \bibinfo {author} {\bibfnamefont {W.}~\bibnamefont {Song}}, \bibinfo
  {author} {\bibfnamefont {Z.}~\bibnamefont {Liu}}, \bibinfo {author}
  {\bibfnamefont {D.}~\bibnamefont {Shen}}, \bibinfo {author} {\bibfnamefont
  {Y.}~\bibnamefont {Huang}}, \bibinfo {author} {\bibfnamefont
  {K.}~\bibnamefont {Liu}}, \bibinfo {author} {\bibfnamefont {H.}~\bibnamefont
  {Lei}},\ and\ \bibinfo {author} {\bibfnamefont {S.}~\bibnamefont {Wang}},\
  }\href@noop {} {\bibfield  {journal} {\bibinfo  {journal} {Phys. Rev. X}\
  }\textbf {\bibinfo {volume} {11}},\ \bibinfo {pages} {041010} (\bibinfo
  {year} {2021})}\BibitemShut {NoStop}%
\bibitem [{\citenamefont {Cho}\ \emph {et~al.}(2021)\citenamefont {Cho},
  \citenamefont {Ma}, \citenamefont {Xia}, \citenamefont {Yang}, \citenamefont
  {Liu}, \citenamefont {Huang}, \citenamefont {Jiang}, \citenamefont {Lu},
  \citenamefont {Liu}, \citenamefont {Liu}, \citenamefont {Li}, \citenamefont
  {Wang}, \citenamefont {Liu}, \citenamefont {Jia}, \citenamefont {Guo},
  \citenamefont {Liu},\ and\ \citenamefont {Shen}}]{ChoPRL2021}%
  \BibitemOpen
  \bibfield  {author} {\bibinfo {author} {\bibfnamefont {S.}~\bibnamefont
  {Cho}}, \bibinfo {author} {\bibfnamefont {H.}~\bibnamefont {Ma}}, \bibinfo
  {author} {\bibfnamefont {W.}~\bibnamefont {Xia}}, \bibinfo {author}
  {\bibfnamefont {Y.}~\bibnamefont {Yang}}, \bibinfo {author} {\bibfnamefont
  {Z.}~\bibnamefont {Liu}}, \bibinfo {author} {\bibfnamefont {Z.}~\bibnamefont
  {Huang}}, \bibinfo {author} {\bibfnamefont {Z.}~\bibnamefont {Jiang}},
  \bibinfo {author} {\bibfnamefont {X.}~\bibnamefont {Lu}}, \bibinfo {author}
  {\bibfnamefont {J.}~\bibnamefont {Liu}}, \bibinfo {author} {\bibfnamefont
  {Z.}~\bibnamefont {Liu}}, \bibinfo {author} {\bibfnamefont {J.}~\bibnamefont
  {Li}}, \bibinfo {author} {\bibfnamefont {J.}~\bibnamefont {Wang}}, \bibinfo
  {author} {\bibfnamefont {Y.}~\bibnamefont {Liu}}, \bibinfo {author}
  {\bibfnamefont {J.}~\bibnamefont {Jia}}, \bibinfo {author} {\bibfnamefont
  {Y.}~\bibnamefont {Guo}}, \bibinfo {author} {\bibfnamefont {J.}~\bibnamefont
  {Liu}},\ and\ \bibinfo {author} {\bibfnamefont {D.}~\bibnamefont {Shen}},\
  }\href@noop {} {\bibfield  {journal} {\bibinfo  {journal} {Phys. Rev. Lett.}\
  }\textbf {\bibinfo {volume} {127}},\ \bibinfo {pages} {236401} (\bibinfo
  {year} {2021})}\BibitemShut {NoStop}%
\bibitem [{\citenamefont {Li}\ \emph {et~al.}(2022{\natexlab{c}})\citenamefont
  {Li}, \citenamefont {Wu}, \citenamefont {Liu}, \citenamefont {Polley},
  \citenamefont {Guo}, \citenamefont {Wang}, \citenamefont {Han}, \citenamefont
  {Dendzik}, \citenamefont {Berntsen}, \citenamefont {Thiagarajan},
  \citenamefont {Shi}, \citenamefont {Schnyder},\ and\ \citenamefont
  {Tjernberg}}]{LiPRR2022}%
  \BibitemOpen
  \bibfield  {author} {\bibinfo {author} {\bibfnamefont {C.}~\bibnamefont
  {Li}}, \bibinfo {author} {\bibfnamefont {X.}~\bibnamefont {Wu}}, \bibinfo
  {author} {\bibfnamefont {H.}~\bibnamefont {Liu}}, \bibinfo {author}
  {\bibfnamefont {C.}~\bibnamefont {Polley}}, \bibinfo {author} {\bibfnamefont
  {Q.}~\bibnamefont {Guo}}, \bibinfo {author} {\bibfnamefont {Y.}~\bibnamefont
  {Wang}}, \bibinfo {author} {\bibfnamefont {X.}~\bibnamefont {Han}}, \bibinfo
  {author} {\bibfnamefont {M.}~\bibnamefont {Dendzik}}, \bibinfo {author}
  {\bibfnamefont {M.~H.}\ \bibnamefont {Berntsen}}, \bibinfo {author}
  {\bibfnamefont {B.}~\bibnamefont {Thiagarajan}}, \bibinfo {author}
  {\bibfnamefont {Y.}~\bibnamefont {Shi}}, \bibinfo {author} {\bibfnamefont
  {A.~P.}\ \bibnamefont {Schnyder}},\ and\ \bibinfo {author} {\bibfnamefont
  {O.}~\bibnamefont {Tjernberg}},\ }\href
  {https://doi.org/10.1103/PhysRevResearch.4.033072} {\bibfield  {journal}
  {\bibinfo  {journal} {Phys. Rev. Res.}\ }\textbf {\bibinfo {volume} {4}},\
  \bibinfo {pages} {033072} (\bibinfo {year} {2022}{\natexlab{c}})}\BibitemShut
  {NoStop}%
\bibitem [{\citenamefont {Kato}\ \emph
  {et~al.}(2022{\natexlab{c}})\citenamefont {Kato}, \citenamefont {Li},
  \citenamefont {Kawakami}, \citenamefont {Liu}, \citenamefont {Nakayama},
  \citenamefont {Wang}, \citenamefont {Moriya}, \citenamefont {Tanaka},
  \citenamefont {Takahashi}, \citenamefont {Yao},\ and\ \citenamefont
  {Sato}}]{KatoCOMMAT2022}%
  \BibitemOpen
  \bibfield  {author} {\bibinfo {author} {\bibfnamefont {T.}~\bibnamefont
  {Kato}}, \bibinfo {author} {\bibfnamefont {Y.}~\bibnamefont {Li}}, \bibinfo
  {author} {\bibfnamefont {T.}~\bibnamefont {Kawakami}}, \bibinfo {author}
  {\bibfnamefont {M.}~\bibnamefont {Liu}}, \bibinfo {author} {\bibfnamefont
  {K.}~\bibnamefont {Nakayama}}, \bibinfo {author} {\bibfnamefont
  {Z.}~\bibnamefont {Wang}}, \bibinfo {author} {\bibfnamefont {A.}~\bibnamefont
  {Moriya}}, \bibinfo {author} {\bibfnamefont {K.}~\bibnamefont {Tanaka}},
  \bibinfo {author} {\bibfnamefont {T.}~\bibnamefont {Takahashi}}, \bibinfo
  {author} {\bibfnamefont {Y.}~\bibnamefont {Yao}},\ and\ \bibinfo {author}
  {\bibfnamefont {T.}~\bibnamefont {Sato}},\ }\href@noop {} {\bibfield
  {journal} {\bibinfo  {journal} {Commun. Mater.}\ }\textbf {\bibinfo {volume}
  {3}},\ \bibinfo {pages} {30} (\bibinfo {year}
  {2022}{\natexlab{c}})}\BibitemShut {NoStop}%
\bibitem [{\citenamefont {Hu}\ \emph {et~al.}(2022{\natexlab{b}})\citenamefont
  {Hu}, \citenamefont {Wu}, \citenamefont {Ortiz}, \citenamefont {Han},
  \citenamefont {Plumb}, \citenamefont {Wilson}, \citenamefont {Schnyder},\
  and\ \citenamefont {Shi}}]{HuPRB2022}%
  \BibitemOpen
  \bibfield  {author} {\bibinfo {author} {\bibfnamefont {Y.}~\bibnamefont
  {Hu}}, \bibinfo {author} {\bibfnamefont {X.}~\bibnamefont {Wu}}, \bibinfo
  {author} {\bibfnamefont {B.~R.}\ \bibnamefont {Ortiz}}, \bibinfo {author}
  {\bibfnamefont {X.}~\bibnamefont {Han}}, \bibinfo {author} {\bibfnamefont
  {N.~C.}\ \bibnamefont {Plumb}}, \bibinfo {author} {\bibfnamefont {S.~D.}\
  \bibnamefont {Wilson}}, \bibinfo {author} {\bibfnamefont {A.~P.}\
  \bibnamefont {Schnyder}},\ and\ \bibinfo {author} {\bibfnamefont
  {M.}~\bibnamefont {Shi}},\ }\href
  {https://doi.org/10.1103/PhysRevB.106.L241106} {\bibfield  {journal}
  {\bibinfo  {journal} {Phys. Rev. B}\ }\textbf {\bibinfo {volume} {106}},\
  \bibinfo {pages} {L241106} (\bibinfo {year}
  {2022}{\natexlab{b}})}\BibitemShut {NoStop}%
\bibitem [{\citenamefont {Xiao}\ \emph {et~al.}()\citenamefont {Xiao},
  \citenamefont {Lin}, \citenamefont {Li}, \citenamefont {Xia}, \citenamefont
  {Zheng}, \citenamefont {Zhang}, \citenamefont {Guo}, \citenamefont {Feng},\
  and\ \citenamefont {Peng}}]{XiaoarXiv2022}%
  \BibitemOpen
  \bibfield  {author} {\bibinfo {author} {\bibfnamefont {Q.}~\bibnamefont
  {Xiao}}, \bibinfo {author} {\bibfnamefont {Y.}~\bibnamefont {Lin}}, \bibinfo
  {author} {\bibfnamefont {Q.}~\bibnamefont {Li}}, \bibinfo {author}
  {\bibfnamefont {W.}~\bibnamefont {Xia}}, \bibinfo {author} {\bibfnamefont
  {X.}~\bibnamefont {Zheng}}, \bibinfo {author} {\bibfnamefont
  {S.}~\bibnamefont {Zhang}}, \bibinfo {author} {\bibfnamefont
  {Y.}~\bibnamefont {Guo}}, \bibinfo {author} {\bibfnamefont {J.}~\bibnamefont
  {Feng}},\ and\ \bibinfo {author} {\bibfnamefont {Y.}~\bibnamefont {Peng}},\
  }\bibinfo {note} {arXiv:2201.05211}\BibitemShut {NoStop}%
\bibitem [{\citenamefont {Li}\ \emph {et~al.}(2022{\natexlab{d}})\citenamefont
  {Li}, \citenamefont {Fabbris}, \citenamefont {Said}, \citenamefont {Sun},
  \citenamefont {Jiang}, \citenamefont {Yin}, \citenamefont {Pai},
  \citenamefont {Yoon}, \citenamefont {Lupini}, \citenamefont {Nelson},
  \citenamefont {Yin}, \citenamefont {Gong}, \citenamefont {Tu}, \citenamefont
  {Lei}, \citenamefont {Cheng}, \citenamefont {Hasan}, \citenamefont {Wang},
  \citenamefont {Yan}, \citenamefont {Thomale}, \citenamefont {Lee},\ and\
  \citenamefont {Miao}}]{LiNCOM2022}%
  \BibitemOpen
  \bibfield  {author} {\bibinfo {author} {\bibfnamefont {H.}~\bibnamefont
  {Li}}, \bibinfo {author} {\bibfnamefont {G.}~\bibnamefont {Fabbris}},
  \bibinfo {author} {\bibfnamefont {A.~H.}\ \bibnamefont {Said}}, \bibinfo
  {author} {\bibfnamefont {J.~P.}\ \bibnamefont {Sun}}, \bibinfo {author}
  {\bibfnamefont {Y.-X.}\ \bibnamefont {Jiang}}, \bibinfo {author}
  {\bibfnamefont {J.-X.}\ \bibnamefont {Yin}}, \bibinfo {author} {\bibfnamefont
  {Y.-Y.}\ \bibnamefont {Pai}}, \bibinfo {author} {\bibfnamefont
  {S.}~\bibnamefont {Yoon}}, \bibinfo {author} {\bibfnamefont {A.~R.}\
  \bibnamefont {Lupini}}, \bibinfo {author} {\bibfnamefont {C.~S.}\
  \bibnamefont {Nelson}}, \bibinfo {author} {\bibfnamefont {Q.~W.}\
  \bibnamefont {Yin}}, \bibinfo {author} {\bibfnamefont {C.~S.}\ \bibnamefont
  {Gong}}, \bibinfo {author} {\bibfnamefont {Z.~J.}\ \bibnamefont {Tu}},
  \bibinfo {author} {\bibfnamefont {H.~C.}\ \bibnamefont {Lei}}, \bibinfo
  {author} {\bibfnamefont {J.-G.}\ \bibnamefont {Cheng}}, \bibinfo {author}
  {\bibfnamefont {M.~Z.}\ \bibnamefont {Hasan}}, \bibinfo {author}
  {\bibfnamefont {Z.}~\bibnamefont {Wang}}, \bibinfo {author} {\bibfnamefont
  {B.}~\bibnamefont {Yan}}, \bibinfo {author} {\bibfnamefont {R.}~\bibnamefont
  {Thomale}}, \bibinfo {author} {\bibfnamefont {H.~N.}\ \bibnamefont {Lee}},\
  and\ \bibinfo {author} {\bibfnamefont {H.}~\bibnamefont {Miao}},\ }\href@noop
  {} {\bibfield  {journal} {\bibinfo  {journal} {Nat. Commun.}\ }\textbf
  {\bibinfo {volume} {13}},\ \bibinfo {pages} {6348} (\bibinfo {year}
  {2022}{\natexlab{d}})}\BibitemShut {NoStop}%
\bibitem [{\citenamefont {Li}\ \emph {et~al.}()\citenamefont {Li},
  \citenamefont {Zhao}, \citenamefont {Ortiz}, \citenamefont {Oey},
  \citenamefont {Wang}, \citenamefont {Wilson},\ and\ \citenamefont
  {Zeljkovic}}]{LiarXiv2022}%
  \BibitemOpen
  \bibfield  {author} {\bibinfo {author} {\bibfnamefont {H.}~\bibnamefont
  {Li}}, \bibinfo {author} {\bibfnamefont {H.}~\bibnamefont {Zhao}}, \bibinfo
  {author} {\bibfnamefont {B.}~\bibnamefont {Ortiz}}, \bibinfo {author}
  {\bibfnamefont {Y.}~\bibnamefont {Oey}}, \bibinfo {author} {\bibfnamefont
  {Z.}~\bibnamefont {Wang}}, \bibinfo {author} {\bibfnamefont {S.~D.}\
  \bibnamefont {Wilson}},\ and\ \bibinfo {author} {\bibfnamefont
  {I.}~\bibnamefont {Zeljkovic}},\ }\bibinfo {note}
  {arXiv:2203.15057}\BibitemShut {NoStop}%
\end{thebibliography}
%apsrev4-2.bst 2019-01-14 (MD) hand-edited version of apsrev4-1.bst
%Control: key (0)
%Control: author (72) initials jnrlst
%Control: editor formatted (1) identically to author
%Control: production of article title (-1) disabled
%Control: page (0) single
%Control: year (1) truncated
%Control: production of eprint (0) enabled
%

\end{document}